\documentclass[twoside,twocolumn,9pt]{article}
\usepackage{extsizes}
\usepackage[super,sort&compress,comma]{natbib} 
\usepackage[version=3]{mhchem}
\usepackage[left=1.5cm, right=1.5cm, top=1.785cm, bottom=2.0cm]{geometry}
\usepackage{balance}
\usepackage{mathptmx}
\usepackage{sectsty}
\usepackage{graphicx} 
\usepackage{lastpage}
\usepackage[format=plain,justification=justified,singlelinecheck=false,font={stretch=1.125,small,sf},labelfont=bf,labelsep=space]{caption}
\usepackage{float}
\usepackage{fancyhdr}
\usepackage{fnpos}
\usepackage[english]{babel}
\addto{\captionsenglish}{%
  
}
\usepackage{array}
\usepackage{droidsans}
\usepackage{charter}
\usepackage[T1]{fontenc}
\usepackage[usenames,dvipsnames]{xcolor}
\usepackage{setspace}
\usepackage[compact]{titlesec}
\usepackage{hyperref}

\usepackage{lineno}
\usepackage{bm, amssymb, color}

\newcommand{\R}{\mathbb{R}}
\newcommand{\fn}[2]{\mathinner{#1\mathopen{\left(#2\right)}}}
\newcommand{\vect}[1]{\bm{#1}}

\newcommand{\phimax}{\phi_\mathrm{max}}

\usepackage[colorinlistoftodos,shadow,textwidth=18mm]{todonotes}

\usepackage{soul}
\setstcolor{red}

\usepackage{epstopdf}

\definecolor{cream}{RGB}{222,217,201}

\begin{document}

\pagestyle{fancy}
\thispagestyle{plain}
\fancypagestyle{plain}{
\renewcommand{\headrulewidth}{0pt}
}

\makeFNbottom
\makeatletter
\renewcommand\LARGE{\@setfontsize\LARGE{15pt}{17}}
\renewcommand\Large{\@setfontsize\Large{12pt}{14}}
\renewcommand\large{\@setfontsize\large{10pt}{12}}
\renewcommand\footnotesize{\@setfontsize\footnotesize{7pt}{10}}
\makeatother

\renewcommand{\thefootnote}{\fnsymbol{footnote}}
\renewcommand\footnoterule{\vspace*{1pt}%
\color{cream}\hrule width 3.5in height 0.4pt \color{black}\vspace*{5pt}} 
\setcounter{secnumdepth}{5}

\makeatletter 
\renewcommand\@biblabel[1]{#1}            
\renewcommand\@makefntext[1]%
{\noindent\makebox[0pt][r]{\@thefnmark\,}#1}
\makeatother 
\renewcommand{\figurename}{\small{Fig.} ~}
\sectionfont{\sffamily\Large}
\subsectionfont{\normalsize}
\subsubsectionfont{\bf}
\setstretch{1.125} 
\setlength{\skip\footins}{0.8cm}
\setlength{\footnotesep}{0.25cm}
\setlength{\jot}{10pt}
\titlespacing*{\section}{0pt}{4pt}{4pt}
\titlespacing*{\subsection}{0pt}{15pt}{1pt}

\fancyfoot{}
\fancyfoot[LO,RE]{\vspace{-7.1pt}\includegraphics[height=9pt]{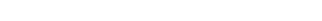}}
\fancyfoot[CO]{\vspace{-7.1pt}\hspace{13.2cm}\includegraphics{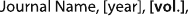}}
\fancyfoot[CE]{\vspace{-7.2pt}\hspace{-14.2cm}\includegraphics{head_foot/RF}}
\fancyfoot[RO]{\footnotesize{\sffamily{1--\pageref{LastPage} ~\textbar  \hspace{2pt}\thepage}}}
\fancyfoot[LE]{\footnotesize{\sffamily{\thepage~\textbar\hspace{3.45cm} 1--\pageref{LastPage}}}}
\fancyhead{}
\renewcommand{\headrulewidth}{0pt} 
\renewcommand{\footrulewidth}{0pt}
\setlength{\arrayrulewidth}{1pt}
\setlength{\columnsep}{6.5mm}
\setlength\bibsep{1pt}

\makeatletter 
\newlength{\figrulesep} 
\setlength{\figrulesep}{0.5\textfloatsep} 

\newcommand{\topfigrule}{\vspace*{-1pt}%
\noindent{\color{cream}\rule[-\figrulesep]{\columnwidth}{1.5pt}} }

\newcommand{\botfigrule}{\vspace*{-2pt}%
\noindent{\color{cream}\rule[\figrulesep]{\columnwidth}{1.5pt}} }

\newcommand{\dblfigrule}{\vspace*{-1pt}%
\noindent{\color{cream}\rule[-\figrulesep]{\textwidth}{1.5pt}} }

\makeatother

\twocolumn[
  \begin{@twocolumnfalse}
{
{\includegraphics[height=55pt]{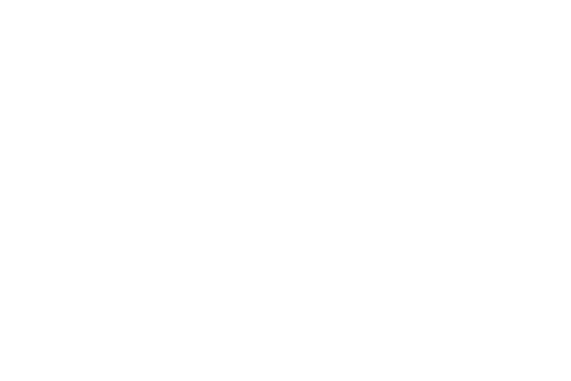}}\\[1ex]
\includegraphics[width=18.5cm]{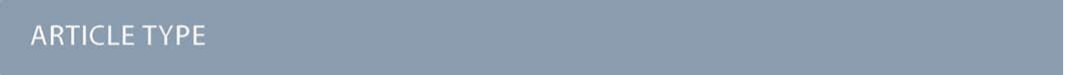}}\par
\vspace{1em}
\sffamily
\begin{tabular}{m{4.5cm} p{13.5cm} }

\includegraphics{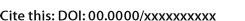} & \noindent\LARGE{\textbf{Existence of Nonequilibrium Glasses in the Degenerate Stealthy Hyperuniform Ground-State Manifold}} \\
\vspace{0.3cm} & \vspace{0.3cm} \\

 & \noindent\large{Salvatore Torquato$^{\ast}$\textit{$^{a,b,c,d}$} and Jaeuk Kim\textit{$^{c,b,a}$}
 } \\

\includegraphics{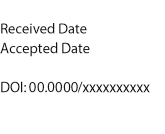} & \noindent\normalsize{
Stealthy interactions are an emerging class of nontrivial, bounded long-ranged oscillatory pair potentials with classical ground states that can be disordered, hyperuniform, and infinitely degenerate. Their hybrid crystal-liquid nature endows them with novel
physical properties with advantages over their crystalline counterparts.
Here, we show the existence of nonequilibrium hard-sphere glasses within this unusual ground-state manifold as the stealthiness
parameter $\chi$ tends to zero that are remarkably configurationally extremely close to hyperuniform 3D maximally random jammed (MRJ) sphere packings. The latter are prototypical glasses since they are maximally disordered, perfectly rigid, and perfectly nonergodic. Our optimization procedure, which leverages the maximum cardinality of the infinite ground-state set, not only guarantees that our packings are hyperuniform with the same structure-factor scaling exponent as the MRJ state, but they share other salient structural attributes, including
a packing fraction of $0.638$, a mean contact number per particle of 6, gap exponent of $0.44(1)$, and pair correlation functions $\fn{g_2}{r}$
and structures factors $\fn{S}{k}$ that are virtually identical to one another for all $r$ and $k$, respectively.
Moreover, we demonstrate that stealthy
hyperuniform packings can be created within the
disordered regime ($0 < \chi <1/2$) with heretofore unattained
maximal packing fractions. As $\chi$ increases from zero, the particles in this family
of disordered packings always form interparticle contacts, albeit with sparser contact networks as $\chi$ increases from zero, resulting in
linear polymer-like chains of contacting particles with increasingly shorter chain lengths. The capacity to generate ultradense stealthy hyperuniform packings for all $\chi$ opens up new materials applications in optics and acoustics.

} 
\end{tabular}

 \end{@twocolumnfalse} \vspace{0.6cm}

  ]

\renewcommand*\rmdefault{bch}\normalfont\upshape
\rmfamily
\section*{}
\vspace{-1cm}


\footnotetext{\textit{$^{a}$~Department of Chemistry, Princeton University, Princeton, New Jersey 08544, USA}}
\footnotetext{\textit{$^{b}$~Department of Physics, Princeton University, Princeton, New Jersey 08544, USA}}
\footnotetext{\textit{$^{c}$~Princeton Materials Institute, Princeton University, Princeton, New Jersey 08544, USA}}
\footnotetext{\textit{$^{d}$~Program in Applied and Computational Mathematics, Princeton University, Princeton, New Jersey 08544, USA}}

\footnotetext[1]{Email: torquato@princeton.edu}



\section{Introduction}  \label{sec:intro}

Disordered hyperuniform many-particle systems\cite{To03a,To18a} are an emerging exotic
class of amorphous states of matter that arise in a variety of contexts
and fields, including the eigenvalues of random matrices,\cite{Dy70}
nontrivial zeros of the Riemann zeta function,\cite{Mon73} maximally random jammed sphere packings, \cite{Do05d,Ma23} avian photoreceptor mosaics,\cite{Ji14} antigen receptors in the immune system,\cite{Ma15} fermionic ground states,\cite{To08b,Wa24c} nonequilibrium phase
transitions,\cite{Re14,He15,He17b,Zh23} active matter,\cite{Le19a} quasicrystals,
\cite{Za09,Og17} distribution of prime numbers,\cite{To19} the
large-scale structure of the universe,\cite{Ph23} soft polymeric materials,\cite{Ch18} quantum spin liquids,\cite{Ch25} and myriads of other examples (see Ref. \citenum{To18a}
and references therein). 
These correlated disordered systems are characterized by a ``hidden order'' due to the unusual
combination of being statistically isotropic without any long-range
order, like ordinary liquids, and yet anomalously suppress infinite-wavelength density
fluctuations, like perfect crystals and quasicrystals. 

A disordered hyperuniform many-particle system in $d$-dimensional Euclidean space $\mathbb{R}^d$
is one in which the structure factor $S(\mathbf{k})$ vanishes as the wavenumber $k\equiv|{\bf{k}}|$ tends to zero.\cite{To03a,To18a}
An important subclass are disordered {\it stealthy} hyperuniform (SHU) systems in which $S(k) = 0$ within a spherical
``exclusion'' region of radius $K$ centered at the origin, i.e., $S(k)=0$ for $0\leq k <K$.
SHU many-particle systems are derived as classical ground states of many-particle systems
of certain nontrivial, bounded long-ranged oscillatory pair potentials.
Remarkably, these hyperuniform ground states can be disordered 
and infinitely degenerate in the thermodynamic limit.\cite{Uc04b,Ba08,To15} The fact that these singular isotropic amorphous states of matter have the character of crystals ($S(k)=0$ from infinite wavelengths to
intermediate wavelengths of order $2\pi/K$) \cite{Zh17a,Gh18,To18a,Sa24} and liquids (statistical isotropy on small length scales)\cite{To18a}
endow them with novel optical, transport, and mechanical properties with advantages over their crystalline counterparts.\cite{Fl09b,Gk17,Fr17,Cas17,To18c,Bi19,Ro19,Roh20,To21a,Ch21,Zh21,Ro21,Kl22,Ta22,Ta23,Me23,Al23,Gr23b,Zh24,Ku24}

\begin{figure}[h]
    \centering
    \includegraphics[width=0.8\linewidth]{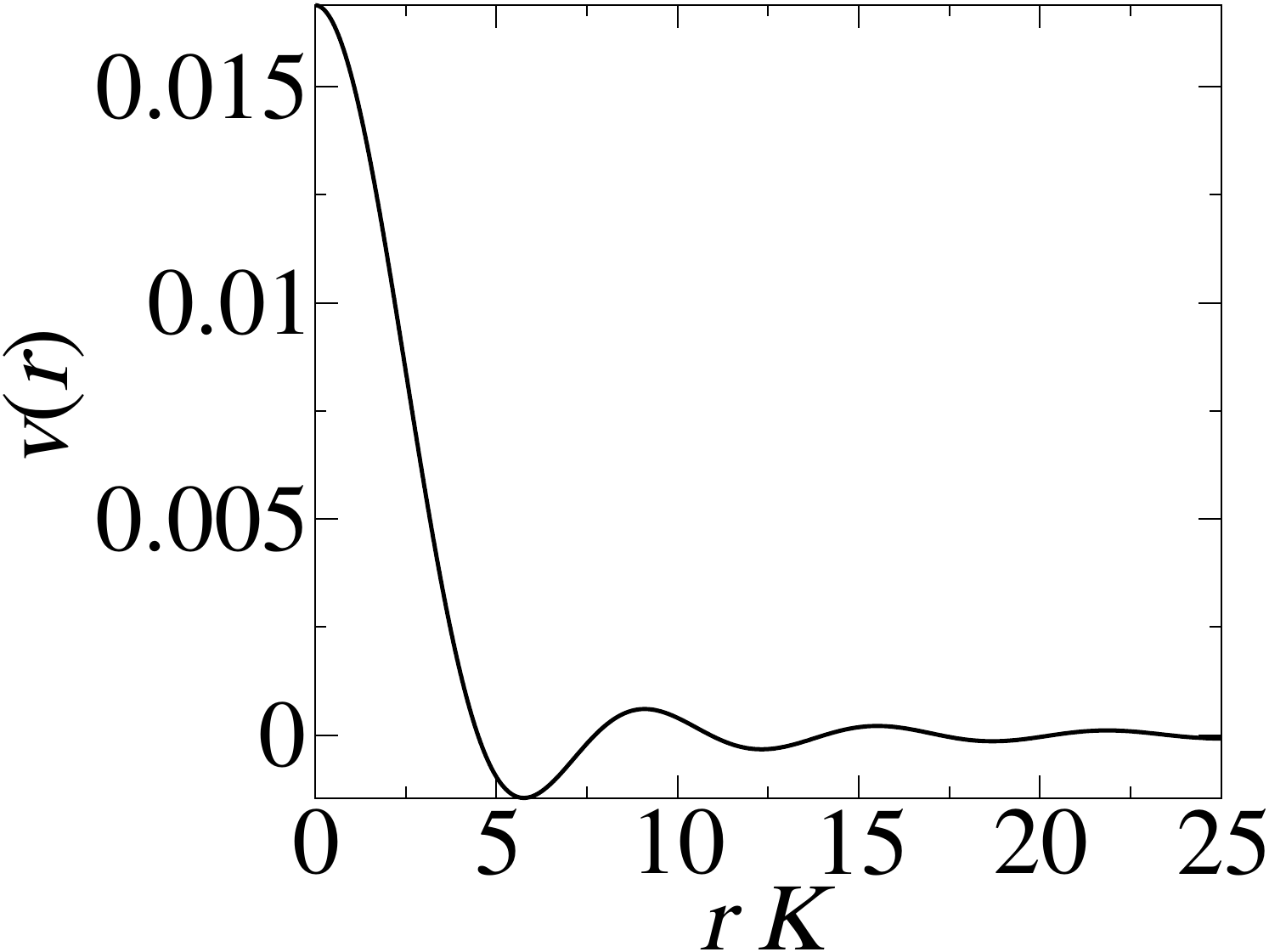}
    \caption{The three-dimensional direct-space long-ranged pair potential $v(r)$ versus the dimensionless distance $rK$ associated with the Fourier transform $\tilde{v}(k)$ that is unity for $k<K$ and zero otherwise, i.e., $\tilde{v}(k)= v_0\Theta(K-k)$.
    We choose $v_0=1$. This pair potential will be used in our
    subsequent computations; see Eq. \eqref{standard-2} with $d=3$.} 
    \label{fig:vr}
\end{figure}

While it is commonplace for quantum-mechanical ground states to be disordered for a variety of typical Hamiltonians,\cite{Wi36,Fe56,Laugh87,To08b,Br20,Se21,Sa21,Kiv23,Wa24c} 
it is unusual for classical many-particle systems to remain disordered down to absolute zero temperature with nontrivial interactions, as is the case for disordered SHU ground states.
The standard collective-coordinate optimization scheme has been used to create disordered SHU ground states
from random initial configurations of $N$ particles within a simulation cell with pair potential $v(r)$ under periodic boundary conditions.\cite{Uc04b,Ba08,Zh15a,Mo23,Sh24}
The total potential energy has the Fourier-space form
\begin{align}
 \Phi(\vect{r}^N) =\frac{\rho}{2 } \sum_{\vect{k} \neq \vect{0}} \fn{\tilde{v}}{\vect{k}}\fn{S}{\vect{k}},
\label{standard}
\end{align}
where ${\tilde v}({\bf k})$ is the Fourier transform of $v(r)$, 
$\rho= N/v_\mathfrak{F}$ is the number density and $v_\mathfrak{F}$ is the volume of the fundamental cell. 
The crucial idea is that if ${\tilde v}({\bf k})$
is bounded and positive with support in the radial
interval $0 \le |{\bf k}| \le K$ and if the particles are rearranged (via optimization) so that the structure factor $S({\bf k})$
is driven to its minimum value of zero for all wavevectors in ``exclusion sphere,'' then 
the system must be at its ground state or global energy minimum. 
This class of functions ${\tilde v}({\bf k})$ lead to
bounded long-ranged oscillatory direct-space
pair potentials $v({\bf r})$, a three-dimensional example of which is shown in Fig.  \ref{fig:vr}.
We note that regardless of the choice of $\tilde{v}({\bf k})$, the functional form of the potential $v({\bf r})$ is always long-ranged, since ${\tilde v}({\bf k})$ is a function that has support only in the exclusion region (i.e., $0\leq |{\bf k}| \leq K$).\cite{To15}

\begin{figure}[h]
    \centering
    \includegraphics[width=0.9\linewidth]{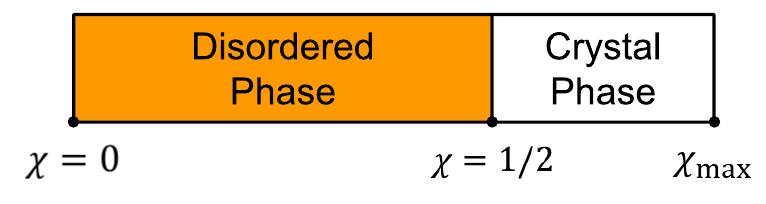}
    \caption{Phase diagram for $d$-dimensional SHU ground states ($T=0$) as a function of $\chi$ for relatively low dimensions with $d \ge 2$.
    The maximal value of $\chi$, denoted by $\chi_{\max}$, depends on the space dimension $d$.
    Adapted from Ref. \citenum{To15}.
    }
    \label{fig:SHU-phase}
\end{figure}

SHU ground states have been numerically generated with ultrahigh accuracy.\cite{Ba08,Zh15a,Mo23}
The disordered ground-state
regime for $d\ge 2$ occurs when $0 < \chi <1/2$,\cite{To15} where $\chi \equiv {M(K)}/{d(N-1)}$ is a
dimensionless stealthiness parameter\footnote{In the thermodynamic limit, $\chi$ is inversely proportional to $\rho$ according to the exact relation\cite{To15}
$\rho \,\chi =v_1(K)/[2d\,(2\pi)^d]$, where $v_1(R)$ 
is the volume of a $d$-dimensional sphere of radius $R$.} that specifies the number of independently constrained wavevectors, $M(K)$,
(within the exclusion sphere of radius $K$) relative to the total number of degrees of freedom of the system, $d(N-1)$.\footnote{Although there are $2M(K)+1$ wavevectors inside the spherical exclusion region of radius $K$ centered at the origin, only $M(K)$ of them are independently constrained because $S({\bf k})$ is inversion symmetric, i.e., $S({\bf k})=S(-{\bf k})$.}
Importantly, the dimensionality
of the configuration space per particle, $d_c$, decreases linearly with $\chi$ as $d_c=d(1-2\chi)$ in the thermodynamic limit, explaining the transition from infinitely degenerate disordered phases (when $\chi <1/2$) to unique crystal structures when $\chi>1/2$;\cite{To15} see Fig. \ref{fig:SHU-phase}.
Thus, $d_c$ is a measure of the {\it cardinality} of the infinitely degenerate ground-state manifold set,  which, importantly, is maximized 
in the limit $\chi\to 0$. We will see that this distinguished limit
plays a central role in the primary results obtained in this
paper.

In the disordered regime, the nature of the energy landscape allows one to find ground states with exquisite precision\cite{Ba08,Zh15a,Mo23}
with a 100\% success rate for relatively large $N$ from random initial conditions.
In the limit $\chi \to 0$, i.e., when the cardinality 
of the infinitely degenerate manifold is maximized, the ground states are ideal-gas-like (Poisson-like),
and thus particle pairs can get arbitrarily close to one another. As $\chi$ increases from zero, the short-range order and minimum pair separation also increase due to an increase in the number of constrained degrees of freedom in a finite-$N$ system.\cite{Ba08,To15,Zh15a}
Such high-$\chi$ disordered ground states 
have been generated to create SHU {\it sphere packings} 
by circumscribing the points by identical nonoverlapping spheres.\cite{Zh16b} However, the success rate
to find allowable packing configurations among all ground states obtained with even moderate values
of the packing fraction $\phi$ ($<0.25)$ falls off rapidly with $N$ and vanishes in the
thermodynamic limit.\cite{Ki25}

Is it possible to exploit the huge degeneracy of the energy landscape to obtain much denser stealthy ground-state packings by biasing the search in configuration space so that such atypical portions
of the manifold are found?
In this article, we show this possible and, by doing so, shed new light on the nature of the infinitely degenerate ground-state manifold
under the action of a {\it generalized} stealthy long-ranged potential (defined in Eq. \eqref{modified}).  We begin by demonstrating
that this manifold, counterintuitively, in the zero-stealthy limit ($\chi\to 0$) when the cardinality is maximized, contains states that are remarkably configurationally
extremely close to the hyperuniform {\it nonequilibrium} MRJ sphere packings that were recently reported by \citealt{Ma23} using the linear programming packing algorithm of Torquato and Jiao.\cite{To10e}\footnote{Such MRJ-like disordered jammed states have been created
by various protocols and systems under periodic boundary conditions, including rapid compression of 
hard spheres \cite{Do05d,To10e,Ch12,Ma23} to their deep mechanically stable local  ``energy'' ($-\phi$)  
minima (inherent structures \cite{St64b,To10e}), rapid quenching of soft spheres at high $T$ to find inherent structures at $T=0$ via conjugate gradient techniques,\cite{Oh02,Ch12} or as a dynamical phase transition via a biased random organization
protocol.\cite{Wi21}
It is well-known that these current packing protocols
lead to disordered jammed packings with similar but not exactly the same structural features, such as rattler concentrations, gap exponents, hyperuniformity, and force distributions. 
For example, the Torquato-Jiao linear programming algorithm \cite{To10e} produces 50\% less rattlers \cite{At13} than that of the modified Lubachevsky-Stillinger algorithm.\cite{Do05c}
Work on rapid quenching of soft spheres \cite{Oh02,Ch12}
typically use only the soft-core potential in (\ref{modified}), i.e., without the stealthy contribution, and hence cannot guarantee hyperuniformity
of the packing.}
 The MRJ packing state is a prototypical nonequilibrium glass,
 since it is the most disordered
packing subject to strict jamming,\footnote{A packing
is strictly jammed if there are no possible collective rearrangements of some finite subset of particles, and no volume-nonincreasing deformation that can be applied to the packing
without violating the impenetrability constraints of the particles.\cite{To10c}} resulting in packings
that are perfectly nonergodic (i.e., permanently trapped in configuration space) and possess infinite elastic moduli.\cite{To00a,To10c}
For this reason, it is remarkable the aforementioned
special ground-state packings and MRJ packings share the
following salient structural attributes: 
pair correlation functions $g_2(r)$
and structures factors $S(k)$ that are virtually identical to one another for all $r$ and $k$, respectively, a packing fraction of $0.638$, a mean contact number per particle of 6 and a ``gap'' exponent of 0.44. Moreover, our 3D ground-state packings become hyperuniform, but appropriately no longer stealthy, in the thermodynamic (infinite-size) limit with the same structure-factor scaling exponent as the MRJ state.
Figuratively, this is akin to finding a nonequilibrium ``needle'' in a ``haystack.''

We also  show that a large family of other stealthy
hyperuniform packings can be created as 
$\chi$ increases up to 1/2 with heretofore unattained maximal packing fractions.
These packings also have interesting structural characteristics; the particles in this family
of disordered packings always form interparticle contacts, albeit with sparser contact networks as $\chi$ increases from zero,  resulting in
linear polymer-like chains of contacting particles that progressively
possess shorter mean chain lengths.
(The reader is referred to  Ref. \citenum{Ki25} for a comprehensive study of SHU packings with and without soft-core repulsions within the disordered regime for $d=1,2,3$.)
This capacity to generate ultradense SHU packings
opens up new applications in optics and acoustics relative to previous studies that used low-density SHU packings.\cite{Gk17,Ro19,Ki20a,To21a,Me23}

The rest of the paper is organized as follows: In Sec. \ref{sec:method}, we describe the modified collective-coordinate optimization scheme and how it is applied to determine the maximal packing fraction of SHU packings with a given value of $\chi$. Here we also describe how to determine whether the corresponding contact networks percolate.
In Sec. \ref{sec:results}, we provide results for the ultradense SHU packings with $\chi$ tending to zero and select positive values of $\chi$ within the disordered regime, including their maximal packing fractions, pair statistics, and mean contact numbers.
Finally, in Sec. \ref{sec:discussion}, we provide concluding remarks and outlook
for future work.

\section{Simulation Method} \label{sec:method}

\subsection{Generation of Dense Disordered SHU Packings}
\label{sec:sim.}

To achieve the desired goal of creating ultradense disordered SHU packings, we modify the total potential energy function (\ref{standard}) by including an additional pairwise soft-core repulsive potential $\fn{u}{r}$
that is bounded, differentiable,  and positive with support in the finite range $0 \le r < \sigma$:\cite{Ki20a,To21a,Sh24} 
\begin{align}
    \fn{\Phi}{\vect{r}^N} =\frac{\rho}{2 } \sum_{\vect{k} \neq \vect{0}} \fn{\tilde{v}}{\vect{k}}\fn{S}{\vect{k}} + \sum_{i <j} \fn{u}{r_{ij}}.
\label{modified}
\end{align}
We choose ${\tilde v}(k)=v_0\Theta(K-k)$, where
$\Theta(x)$ is the Heaviside step function, yielding the bounded long-ranged pair potential
\begin{equation}
\frac{v(r)}{v_0}= \left(\frac{K}{2 \pi r}\right)^{d/2}J_{d/2}(Kr),
\label{standard-2}
\end{equation}
where $J_\nu (x)$ is the Bessel function of the first kind of order $\nu$. 
The soft-core contribution to the energy (\ref{modified}), embodied by $u(r)$, competes with
standard stealthy contribution, and hence
the modified potential biases the optimization to sample portions of the SHU ground-state manifold
in which the minimum particle pair separation is substantially greater
than those obtained via the stealthy contribution only.
Since both sums in (\ref{modified}) are nonnegative, its
ground states with  $\fn{\Phi}{\vect{r}^N}= 0$ must satisfy the SHU condition
[i.e., $S(k)=0$ for $0\leq k <K$] such that all particle pairs are separated by at
least a {\it targeted} distance $\sigma$.
Then, such ground states 
are mapped to SHU packings with a packing fraction at
least given by $\phi=\rho v_1(\sigma/2)$, where $v_1(a) = \pi^{d/2} a^d/\Gamma(1+d/2)$
is the volume of a $d$-dimensional sphere of radius $a$.
At a specific value of $\chi$, the packing fraction $\phi$ can be increased to its maximum value, $\phimax(\chi)$,  beyond which the ground state ceases to exist, which can be viewed as a  satisfiable-unsatisfiable (SAT-UNSAT) transition.\cite{Ma93,Fr17b}
We numerically determine the values of $\phimax(\chi)$, as described in Sec. \ref{sec:phi_max}. 
The modified potential (\ref{modified}) was previously used to generate the SHU ground-state
sphere packings but only at relatively small packing fractions for optical and elastodynamic applications.\cite{Ki20a,To21a}

A general form of the soft-core repulsion $u(r)$ in the potential energy \eqref{modified} is 
\begin{align}  \label{eq:u}
    \frac{\fn{u}{r}}{\epsilon_0} = \left(1-\frac{r}{\sigma}\right)^\beta \fn{\Theta}{\sigma-r}\quad (\beta>1).
\end{align}
In this work, we set $\beta=2$, yielding the harmonic contact potential,\cite{Oh03,Ch12,Ji21} because it is computationally efficient to implement.

Importantly, the ground states of potential \eqref{modified} are independent of the ratio of $v_0/\varepsilon_0$ if $v_0$ and $\varepsilon_0$ are positive and finite.
We consider a cubic fundamental cell, and set the energy scales to be $v_0 = 1$ and $\epsilon_0 = 100 v_0$, enabling us to generate valid packings that are ground states in a reasonable amount of computational time for $N \approx 400-10774$ and $n_c=10^2-10^4$, where $n_c$ is the number of configurations.
Specifically, using an Intel(R) Xeon(R) CPU (E5-2680, 2.40 GHz), it takes approximately 15 core-hours to generate one 3D ground state with $\chi=0.45$, $\phi = 0.47$, and $N=4000$.
It takes roughly 130 core-hours on the same CPU to generate one 3D ground state with $\chi=9.28\times 10^{-5}$, $N=10774$, and $\phi =0.638$.

The algorithm to find ultradense SHU ground-state packings is performed as follows:
For given parameters of system size $N$, $\chi (>0)$, and target packing fraction $\phi=\phimax(\chi)$, we begin with a random initial condition in a simple cubic fundamental cell and minimize its potential energy $\Phi$ given in Eq. (3) via the low-storage Broyden-Fletcher-Goldfarb-Shanno (L-BFGS) algorithm.\cite{No80,Liu89}
The minimization stops when (i) $\Phi < 5 \times 10^{-20}$,\footnote{This threshold is close to zero energy of the potential, given in Eq. (3), within the double precision of the machine} (ii) the number of evaluations exceeds $5\times 10^6$, or (iii) the mean particle displacements are less than $10^{-15}\rho^{-1/d}$.
After terminating the minimization, we retain the ground-state point pattern that satisfies condition (i) and discard it otherwise.
The values of system size $N$, the number of configurations $n_c$, and stealthiness parameter $\chi$ are listed in Table \ref{tab:nc}.

\subsection{Determination of Maximum Packing Fractions}
\label{sec:phi_max}

We first determine the maximal target packing fraction of 3D SHU ground states with the smallest value of $\chi$, denoted by $\chi_{\min}$.
For the simple cubic fundamental cell we consider, $\chi_{\min}$ is a function of particle number $N$: 
\begin{align}\label{eq:min_chi}
    \chi_\mathrm{min}(N) = \frac{1}{N-1},
\end{align}
where we have used the definition of $\chi$, the fact that the smallest number of constraints is $M(K)=3$ with $K=2\pi/L$, and $L$ is the side length of the fundamental cell.
The explicit formula for the total potential with $\chi=\chi_{\min}(N)$ is provided in Eq. \eqref{eq:zero-limit} in Appendix \ref{app:Phi_0}.
Thus, $\chi_{\min}$ vanishes as $N$ tends to infinity.
At each system size $N$, we numerically determine $\phimax(\chi_{\min})$ as follows:
For a target packing fraction $\phi$, we carry out the optimization mentioned above (Sec. \ref{sec:sim.}) for either a given number of initial conditions $n_i(=100)$ or at most 1680 core-hours, equivalent to one week for 10 threads of an Intel(R) Xeon(R) CPU (E5-2680, 2.40 GHz). The success ratio is obtained by dividing the number of ground states by the number of used initial conditions.
This search is done by varying $\phi$ from 0.630 to 0.640 in increments of 0.001.
The values of $\phimax(\chi)$ are determined as the largest $\phi$ such that the success ratio is at least 7\%.
Using the determined value of $\phimax(\chi)$ (see Table \ref{tab:gap}), we then generate $n_c$ independent ground states (as tabulated in Table \ref{tab:nc}).

Using the same procedure, we also numerically determine $\phimax(\chi)$ for $\chi=0.10,0.20,0.30,0.35,0.40,$ and $0.45$, but apply a slightly looser criterion
than for $\chi_{min}$ to save overall computation cost 
in both determination of $\phimax(\chi)$ and generation of the ultradense SHU packings.
At each $\chi$ value, we use either a given number of initial conditions $n_i(=50)$ or, at most, 840 core hours to obtain ground states.
We consider the largest value of $\phi$ achieved at least 10\% of success ratio as $\phimax(\chi)$.
This search is done by varying $\phi$ from 0.20 to the packing fraction of the densest lattice packing in increments of 0.01.
The obtained values of $\phimax(\chi)$ are tabulated in Table \ref{tab:perc}.

\subsection{Percolation of contact networks}
\label{sec:percolation}

For the contact networks explained in the caption of Fig. \ref{images}, we determine whether they percolate by a simplified version of the algorithm for general networks, explained in Refs. \citenum{Ne00, Zh16b}.
Specifically, we begin by creating a contact network of a configuration in which each vertex (or node) is a particle and an edge is formed between any two particles whose separation is less than $\sigma + 10^{-3}\rho^{-1/3}$. 
For such a network, we randomly choose an initial vertex and recursively search for all vertices connected to it. 
This search stops under two conditions: (i) when there are no further vertices to explore or (ii) when a newly searched vertex is a periodic copy of the initial vertex.
When condition (ii) is met, we terminate the search for that configuration and consider the associated contact network to be percolated.
Otherwise, we repeat this procedure up to 1000 different initial vertices for each configuration.
Table \ref{tab:perc} of the Results section provides the percentage of contact
networks that percolate, i.e., topologically connected
across the sample.

\section{Results}\label{sec:results}

Employing our modified collective-coordinate optimization procedure described in Sec. \ref{sec:method}, we demonstrate that our disordered packings,
as $\chi$ tends to zero, become hyperuniform and {\it non-stealthy} in the thermodynamic limit with the same structure-factor scaling exponent as the MRJ state reported in Ref. \citenum{Ma23}.
Furthermore, they share other salient structural attributes, including a packing fraction of $0.638$, a mean contact number per particle of $6$ (isostaticity), a gap exponent $\gamma=0.44$, and pair correlation functions $g_2(r)$ and structure factors $S(k)$ that are virtually identical to one another for all $r$ and $k$, respectively. 
We begin by showing that we can create other stealthy
hyperuniform packings whose particles always form interparticle contacts
for positive values of $\chi$ as it increases up to $1/2$ with heretofore unattained high packing fractions and determine their 
maximal packing fractions as well as mean contact numbers per particle. 
Subsequently, we focus on the pair statistics of the hyperuniform packings in the limit $\chi\to 0$ and SHU packings with $\chi=0.45$.
We determine the limit $\chi\to0$\; by applying $\chi=\chi_{\min}(N)$ given by Eq. \eqref{eq:min_chi} and the associated total potential $\Phi$ given in Eq. \eqref{eq:zero-limit} for a
given $N$ to a sequence of increasing system sizes with $N=400,~1198,~3592$ and $10774$, yielding $\chi_{\min}=2.51 \times 10^{-3},  8.35  \times 10^{-4},  2.78  \times 10^{-4}$ and  $9.27\times 10^{-5}$, respectively, and then extrapolating to the limit $\chi=0$ by increasing $N$.
Importantly, the potential \eqref{eq:zero-limit} guarantees the obtained ground states are (stelathy) hyperuniform for any $N$, different from compression algorithms.
Henceforth, we will refer to this limit simply as $\chi=0^+$.

\subsection{Maximal packing fractions and mean contact numbers}

\begin{table}[h]
\caption{Maximal packing fractions $\phimax(0^+)$ and fitted values of the gap exponent $\gamma$ for our ultradense SHU ground-state packings with $\chi=0^+$, and 3D MRJ packings. 
 Data for the MRJ packings were obtained in Ref. \citenum{Ma23}.
    The values of $\chi_{\min}$ are given in Eq. \eqref{eq:min_chi}.
 	The values in parentheses stand for the statistical errors.
 	\label{tab:gap}
 	}
 	\begin{tabular}{c r c c}
 	\hline 
 	Models &	$N$	& $\phimax(0^+)$& $\gamma$ \\
 	\hline
 	SHU packings &	400	& 0.636	&0.45(1)	\\
 	SHU packings & 1198& 0.637&	0.44(1)	\\
 	SHU packings & 3592& 0.637&	0.44(1)	\\
 	SHU packings & 10774& 0.638&	0.44(1)	\\
 	\hline 		
 	3D MRJ packing 	&	5000& 0.638	&0.44(1)	\\
 	\hline
 	\end{tabular}
\end{table}

\begin{figure}[h]
\centering
\includegraphics[width=0.45\textwidth,keepaspectratio,clip=]{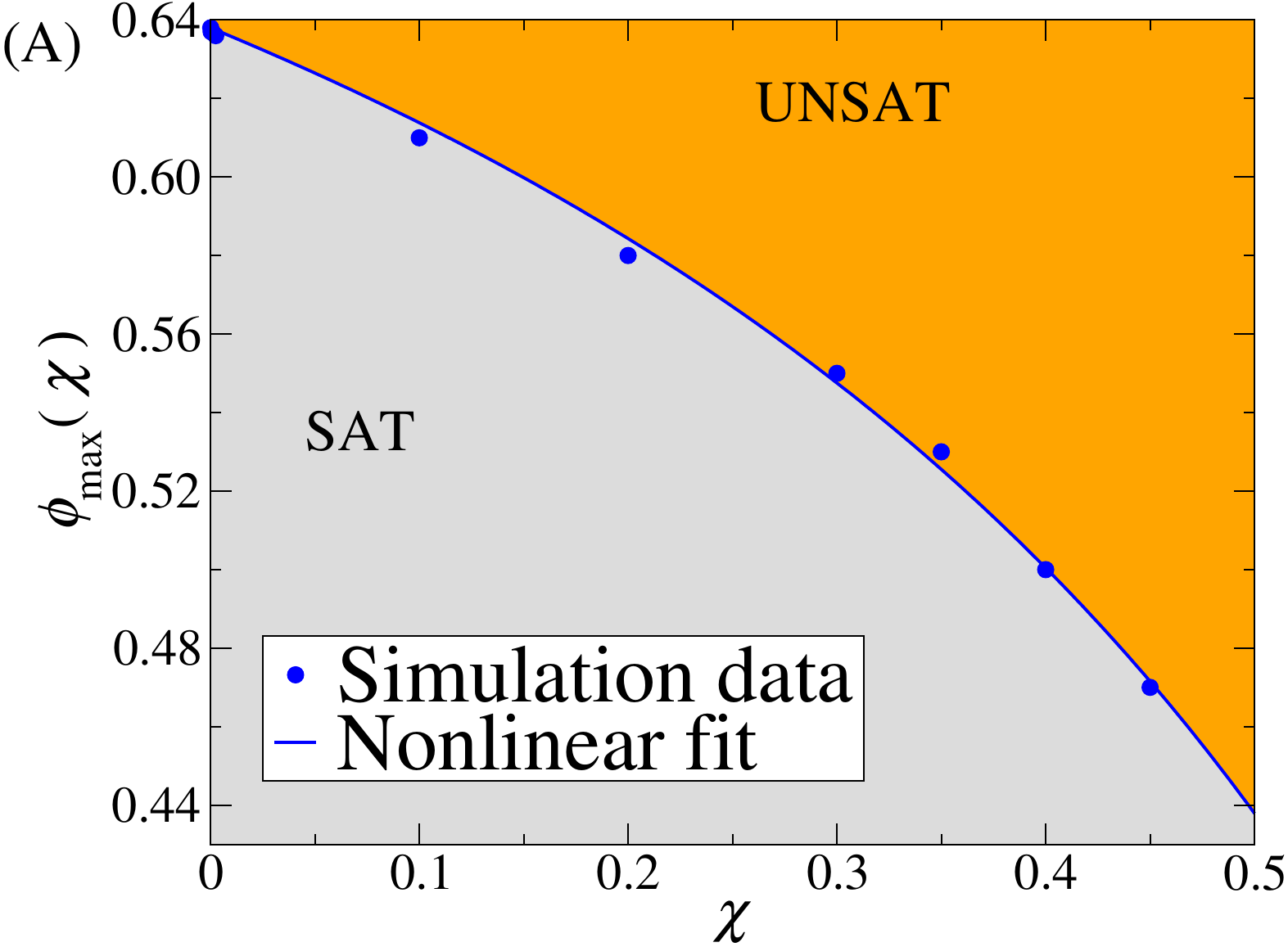}

\includegraphics[width=0.45\textwidth,keepaspectratio,clip=]{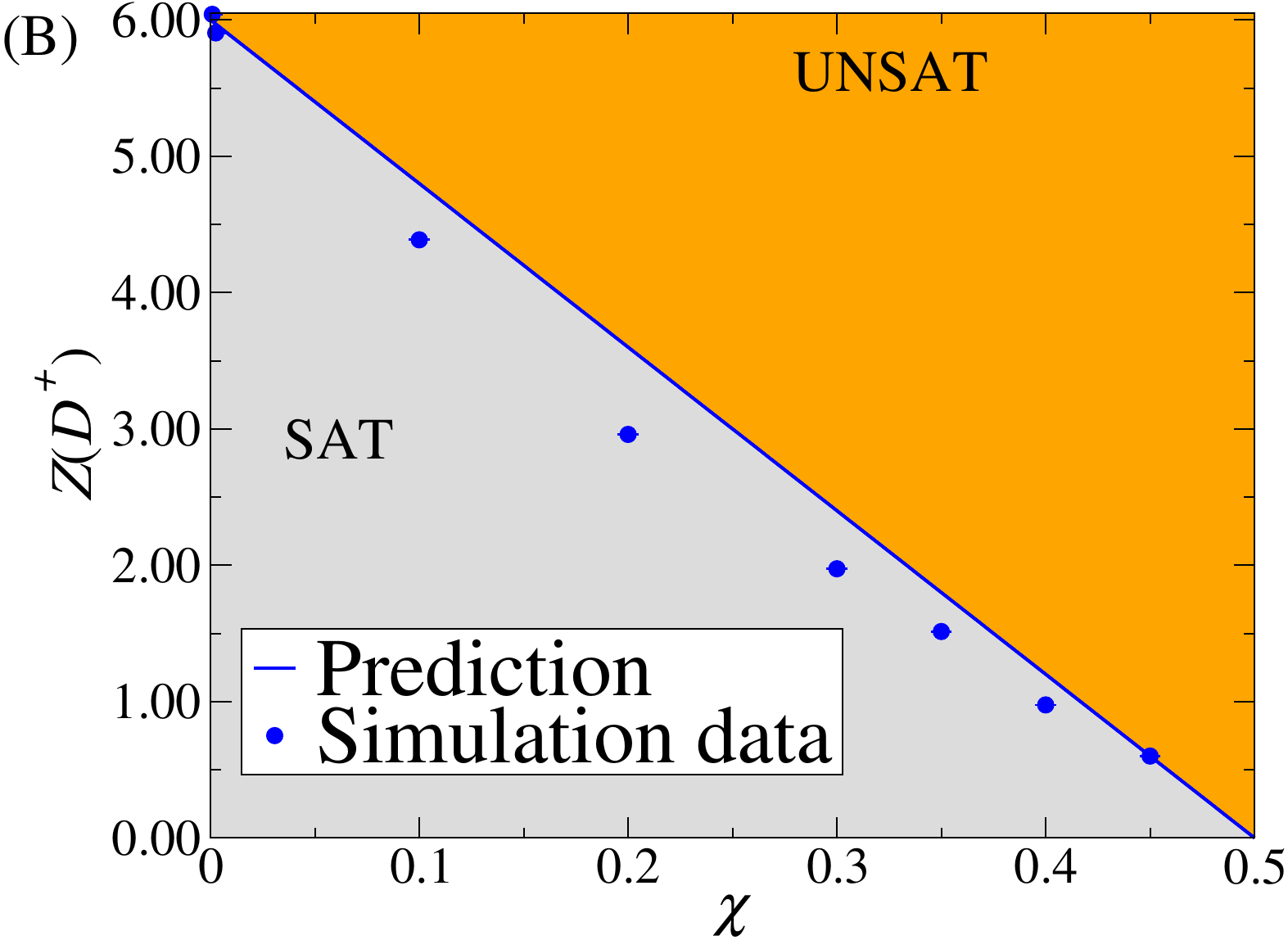}
\caption{ Parameters of 3D disordered ultradense SHU sphere-packing ground states across stealthiness parameter $\chi$.
(A) Maximal packing fraction $\phimax(\chi)$ as a function of $\chi$.
The filled circles are simulation data (with $N$ up to 10774 and $10^2-10^4$ configurations) and the solid curve
is the [1,1] Pad{\'e} approximant of the data given by Eq. \eqref{eq:max-phi_soft}.
(B) Mean contact number per particle ${Z}(D^+)$ as a function of $\chi$.
The solid line is the theoretical prediction \eqref{eq:Z-chi}.
The data in both (A) and (B) are tabulated in Table \ref{tab:perc}
in the Results section.
Note that the gray and orange regions in (A) and (B) correspond to the SAT and UNSAT phases\cite{Ma93,Fr17b} of the ground state of Eq. \eqref{modified}, respectively.
The reader is referred to Ref. \citenum{Ki25} for details.
\label{max-phi}
}
\end{figure}

\begin{figure*}[hbt]
\centerline{\includegraphics[width=0.81\textwidth,keepaspectratio,clip=]{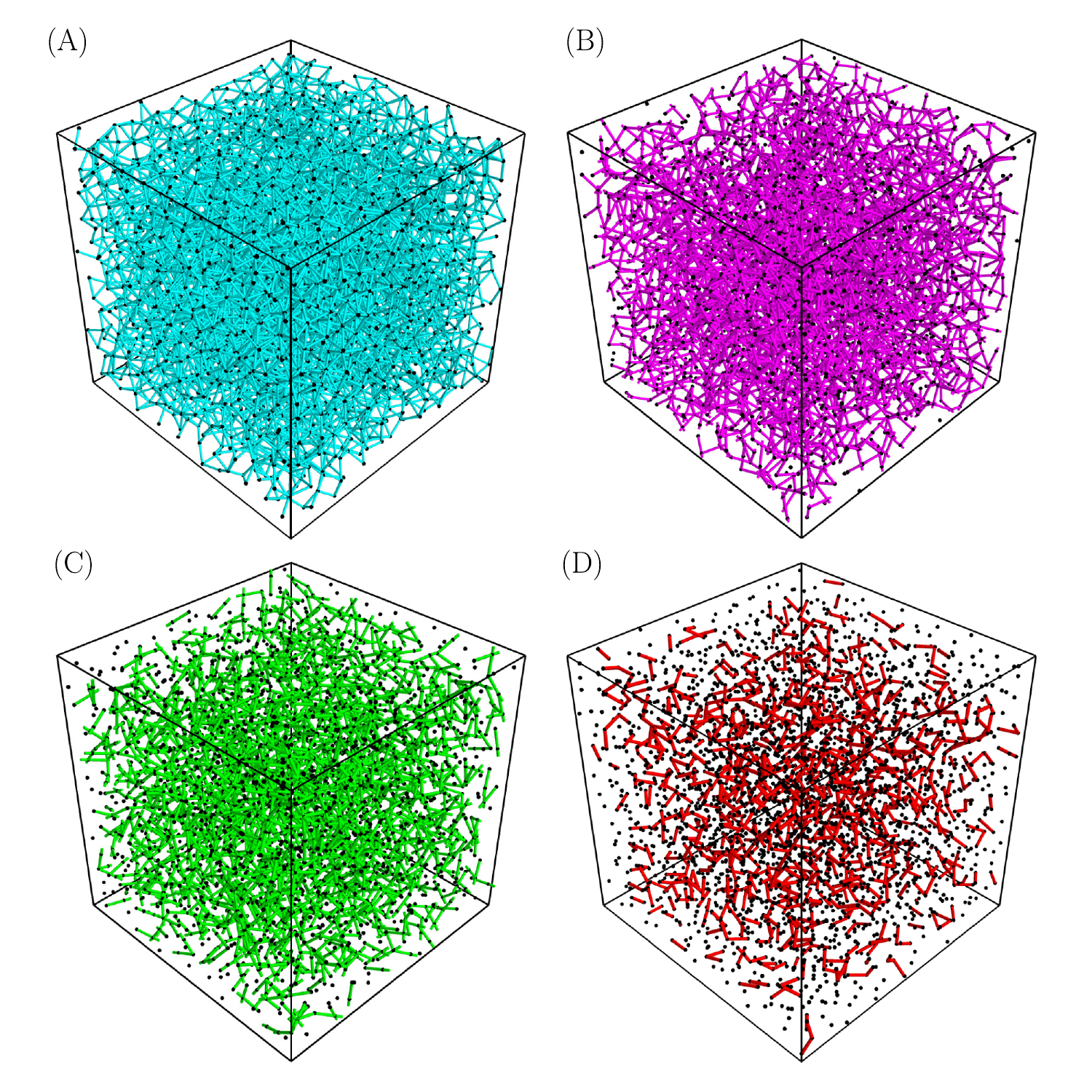}}
\caption{Contact networks of 3D SHU sphere packings for various $\chi$ values: (A) $\chi=\chi_{\min}(N)$, $N=3592$, $\phi=0.637$ and $Z(D^{+}) \approx  6.0 $; 
(B) $\chi=0.20$, $N=4000$, $\phi=0.58$, and $Z(D^+)\approx 3.0$; 
(C) $\chi=0.35$, $N=4000$, $\phi=0.53$, and $Z(D^+)\approx 1.5$;
(D) $\chi=0.45$, $N=4000$, $\phi=0.47$ and $Z(D^{+}) \approx  0.60 $.~
Here, the sizes of the cubic frames are identical.
The centroids of the particles are depicted as black dots in all panels.  
In each panel, the bonds are drawn between particles with colored lines if pairs of particle centers are closer than $\sigma+0.001\rho^{-1/3}$.
}
\label{images}
\end{figure*}

Application of our modified collective-coordinate optimization procedure (see Sec. \ref{sec:method} for details) within the disordered regime ($0 < \chi <1/2$)
yields
achievable packing fractions that are substantially larger than those 
that can be obtained from the standard collective-coordinate
procedure without the soft-core repulsive potential. Importantly, the maximal packing
fraction decreases (not increases) as $\chi$ increases and approaches $\chi=1/2$, as seen in Table \ref{tab:perc} and Fig. \ref{max-phi}(A).
The fact that
$\phi_{\max}$ is highest in the small-$\chi$ limit and then monotonically decreases with  $\chi$ up to $\chi=1/2$ is consistent with the decrease in the relative degrees of freedom as $\chi$ increases.
Such $\chi$-dependence on $\phimax(\chi)$ can be well approximated by the following [1,1] Pad\'e approximant:
\begin{align} \label{eq:max-phi_soft}
\fn{\phimax}{\chi} = 0.638 \frac{1-1.245 \chi}{1- 0.900\chi},
\end{align}
which we see is in very good agreement with the data plotted
in Fig. \ref{max-phi}(A).

It is noteworthy that for all $\chi$ values
within the disordered regime, the particles in this family of packings always form interparticle contacts. The quantity
\begin{equation}
Z(r)=4\pi\rho\int_0^\infty r^2 g_2(r) dr
\end{equation}
is the {\it cumulative coordination number} from which
we can extract the mean contact number per particle, $Z(D^+)$,
and we take $D^+$ (indicating the limit of $r$ to $D$ from above) to be $\sigma_{\max} + 0.001 \rho^{-1/3}$.
As shown in both Fig. \ref{max-phi}(B) and Table \ref{tab:perc},
the mean contact number per particle  is the largest
as $\chi$ tends to zero, where $Z(D^{+})$ achieves the
{\it isostatic} (marginally mechanically stable) value 
of 6 associated with strictly jammed packings 
and then monotonically decreases with $\chi$, implying
{\it sub-isostatic} packings ($Z(D^+) < 6$) that cannot be strictly jammed.\cite{To10c}
Such a decrease in $Z(D^{+})$ with $\chi$ is due to an interplay between two competing constraints in numerical optimizations.
Specifically, for an $N$-particle SHU ground-state packing in $\mathbb{R}^{d}$, $2\chi \times d(N-1)$ degrees of freedom are already constrained among $d(N-1)$ total degrees of freedom by the stealthy hyperuniform condition.
Thus, the remaining degrees of freedom determine the maximum number of constraints from effectively contacting particles:
\begin{align}	\label{eq:Z-chi}
	\fn{Z}{D^+} = 2d (1-2\chi).
\end{align}
This prediction shows good agreement with the simulation data; see Fig. \ref{max-phi}(B).

\begin{table}[h]
	\caption{Summary of parameters of 3D disordered sphere-packing ground states across stealthiness parameter $\chi$.
	Three parameters include maximal packing fractions $\phimax(\chi)$, mean contact number per particle $Z(D^+)$, and the fraction of configurations that have percolated contact networks $p$.
    For other values of $\chi$, we consider $1000$ ground states with $N=4000$ taken from Ref. \citenum{Ki25}.
	The values in the parentheses for $Z(D^+)$ indicate statistical errors.
	\label{tab:perc}
	}
	\begin{tabular}{l l l c}
	\hline 
	$\chi$ &	$\phimax(\chi)$ &	$Z(D^+)$ & $p$ [\%] \\
	\hline
	$0^+$&   0.638&	6.0(1) &	100.0 \\
	0.10&	0.61&		4.388(4)&	100.0\\
	0.20&	0.58&		2.960(4)&	100.0\\
	0.30&	0.55&		1.975(3)&	100.0\\
	0.35&	0.53&		1.514(3)&	~~38.6\\
	0.40&	0.50&		0.975(23)&	~~~~0.0\\
	0.45&	0.47&		0.600(21)&	~~~~0.0\\
	\hline
	\end{tabular}
\end{table}

Figure \ref{images} shows representative contact networks of the ultradense SHU packings at selected
values of $\chi$ between 0 to $1/2$, some of which percolate across the entire system.
At $\chi=0^+$,
shown in Fig. \ref{images}(A), the contact network percolates and effectively satisfies the isostatic condition $Z(D^{+}) \approx 6.0$, implying that these packings are effectively jammed.
As $\chi$ increases from zero, the contact networks
are always sub-isostatic, as predicted by Eq. \eqref{eq:Z-chi}.
For $\chi=0.20$, the contact network percolates, and almost all of the spheres are part of it, but the network becomes sparser (contact number decreases) from the isostatic value of 6 for $\chi\to 0$ with $Z(D^+)\approx 3.0$, as seen in Fig. \ref{images}(B) and Table \ref{tab:perc}.
As $\chi$ increases from $0.20$ to $1/2$, $Z(D^+)$ continues to decrease, i.e., the contact networks
become even sparser, resulting in
{\it linear polymer-like chains} of contacting particles that progressively possess shorter mean chain lengths.
This trend is evident in Fig. \ref{images}(C) for $\chi=0.35$ with $Z(D^+)\approx 1.5$, and in Fig. \ref{images}(D) for $\chi=0.45$ with $Z(D^+)\approx 0.6$.
In addition, approximately 38.6 \% of the contact networks percolate when $\chi=0.35$, and they then cease to percolate for the higher values of $\chi$; see Table \ref{tab:perc} and Sec. \ref{sec:percolation} for details about how we estimate percolation behaviors.
When $\chi=0.45$, the non-percolating contact network consists 
of dimer polymer-like chains. Here, about half of the particles
in the packing are monomers.
The reader is referred to Ref. \citenum{Ki25} for the details behind these results for all $\chi$ 
as well as corresponding findings concerning local spatial statistics (nearest-neighbor and minimal pair-distance
distributions)
for not only 3D cases, but 1D and 2D instances.

\begin{figure}[hbt]
\centerline{\includegraphics[width=0.45\textwidth,keepaspectratio,clip=]{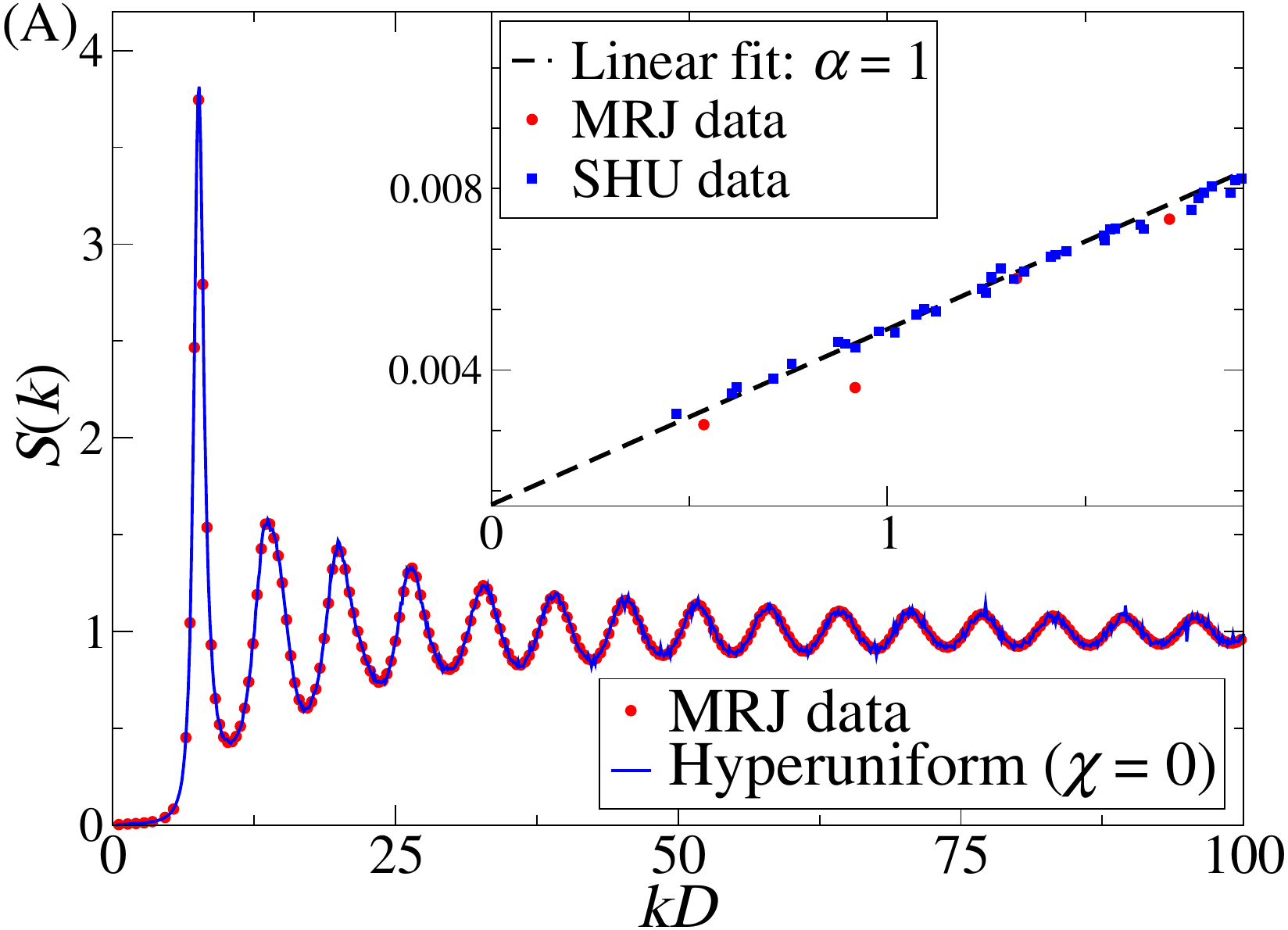}}

\centerline{\includegraphics[width=0.45\textwidth,keepaspectratio,clip=]{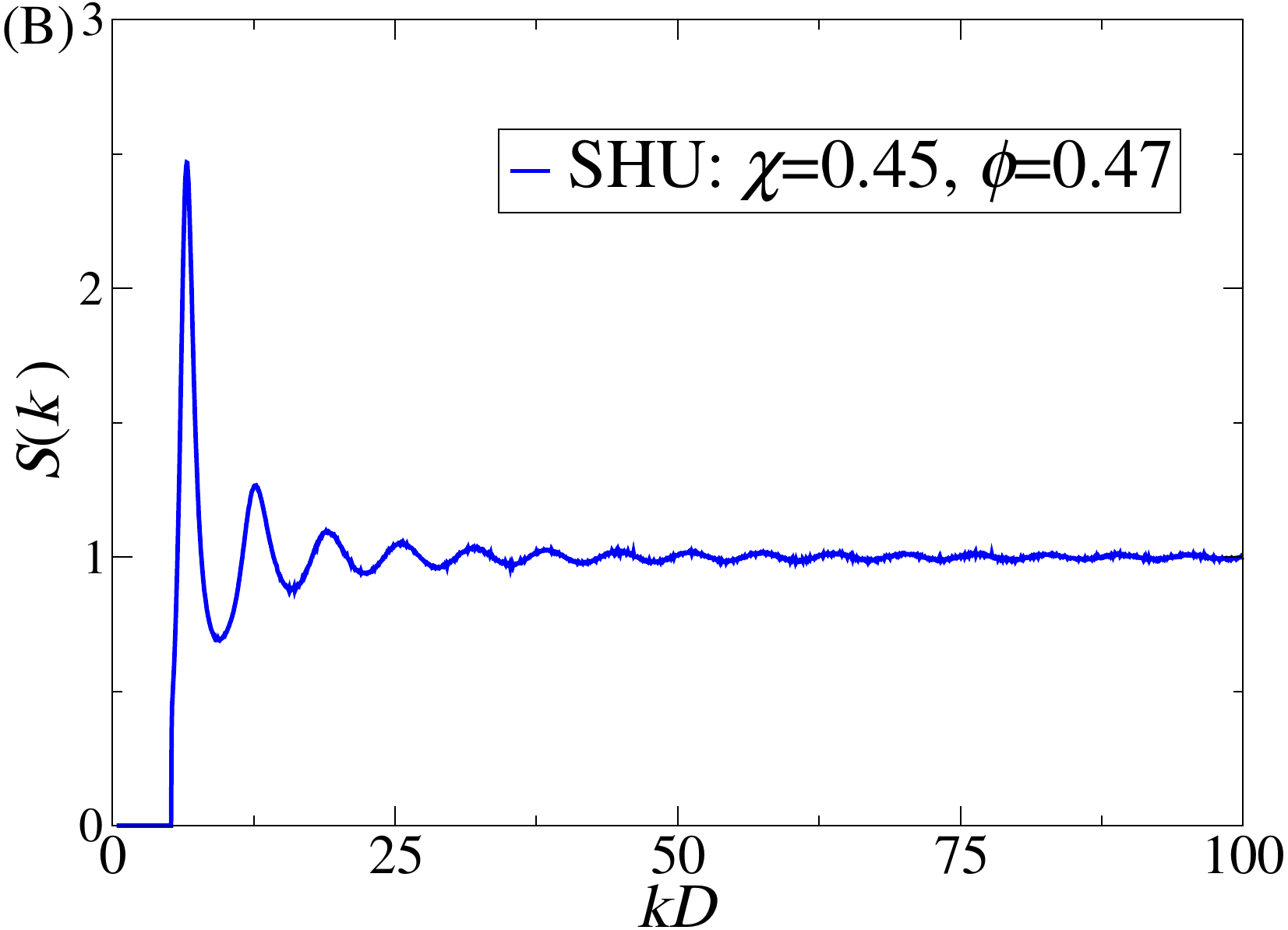}}
\caption{The structure factor $S(k)$ versus dimensionless wavenumber $kD$ of 3D SHU sphere-packing ground states.
(A) Comparison of the data of 3D hyperuniform MRJ sphere packings from Ref. \citenum{Ma23} to the corresponding structure factor for our 3D hyperuniform sphere-packing ground states in the limit $\chi\to0$, as extrapolated from the various system sizes studied, namely, $N=400,~1198,~3592,~10774$. Both models have the same packing fraction, $\phi=0.638$.
The inset shows $S(k)$ in the range of $kD<2$ for SHU sphere-packing ground states with $\chi=\chi_{\min}(N)$ as obtained for the aforementioned sizes and MRJ states.
The black dashed line is a linear fit of the data, i.e., $S(k)\sim k$ in the limit $k\to0$.
(B) 3D SHU sphere-packing ground states with $\chi=0.45$, $N=4000$, $\phi=0.47$, and $n_c=500$ are shown. In both (A) and (B), $D$ is the hard-sphere diameter.
}
\label{3D-Sk}
\end{figure}

\subsection{Structure factors}

An isostatic value of $Z(D^{+}) = 6.0 $ and packing fraction $\phi =0.638$ are not sufficient
to define an MRJ packing. 
Therefore, we now show that the ground-state packings in the limit $\chi\to 0$ (i.e., when the cardinality of the infinitely
degenerate ground-state manifold)
are configurationally extremely close to the MRJ state, as measured by 
structure factors $S(k)$ that are
virtually the same as one another for all $k$.
While MRJ packings are hyperuniform, they are not perfectly stealthy,\cite{Ma23}
and hence the  limit $\chi\to 0$ (corresponding to maximal cardinality of the infinite set of degenerate ground states)
is a necessary condition to check  if SHU
packings have any correspondence to the MRJ state, since the stealthiness vanishes
in this limit while being hyperuniform, which importantly implies the {\it thermodynamic} (infinite-$N$)
limit. 
To approach the large-$N$ limit, we study a sequence of increasing
systems sizes with $N = 400, 1198, 3592$ and $10774$, while taking $\chi=\chi_{\min}(N)$ given by Eq. \eqref{eq:min_chi} and then extrapolate to the limit $N\to \infty$.
We find that as $\chi$ becomes small, the already small stealthy region diminishes in size, and remarkably, the ground states become hyperuniform but not stealthy in the limit $\chi\to0$.
We also confirm that the ultradense SHU packings with $\chi=\chi_{\min}$ exhibit the power-law scaling form $S(k)\sim k$ as $k\to 0$ by extrapolating $S(k)$ in the range of $KD < kD \lesssim 1.3$, which is consistent with those for MRJ state reported in Ref. \citenum{Ma23}; see the inset of Fig. \ref{3D-Sk}(A).
Figure \ref{3D-Sk}(A) also compares the structure factor for a wide range of wavenumbers for the SHU packings at the largest
value of $N$ ($=10774$) to the recent corresponding MRJ data,\cite{Ma23} revealing
that they are virtually identical to one another on the scale of this figure.

By contrast, the structure factors $S(k)$ of the SHU sphere-packing ground states with the larger values of $\chi$ are quite distinct from those in the zero-$\chi$ limit; see Fig. \ref{3D-Sk}(B) for the case $\chi=0.45$.
When the ground states have large values of $\chi$, $S(k)$ clearly have a wide exclusion region in which $S(k)=0$ for small $k$.
Furthermore, the oscillations in $S(k)$ decay more rapidly as wavenumber $k$ increases, reflecting a decrease in the degree of order at short-range length scales due to the formation of dimer polymer-like chains.

\subsection {Pair correlation functions and cumulative coordination number}

\begin{figure}[h!bt]
\centering
\includegraphics[width=0.45\textwidth,keepaspectratio,clip=]{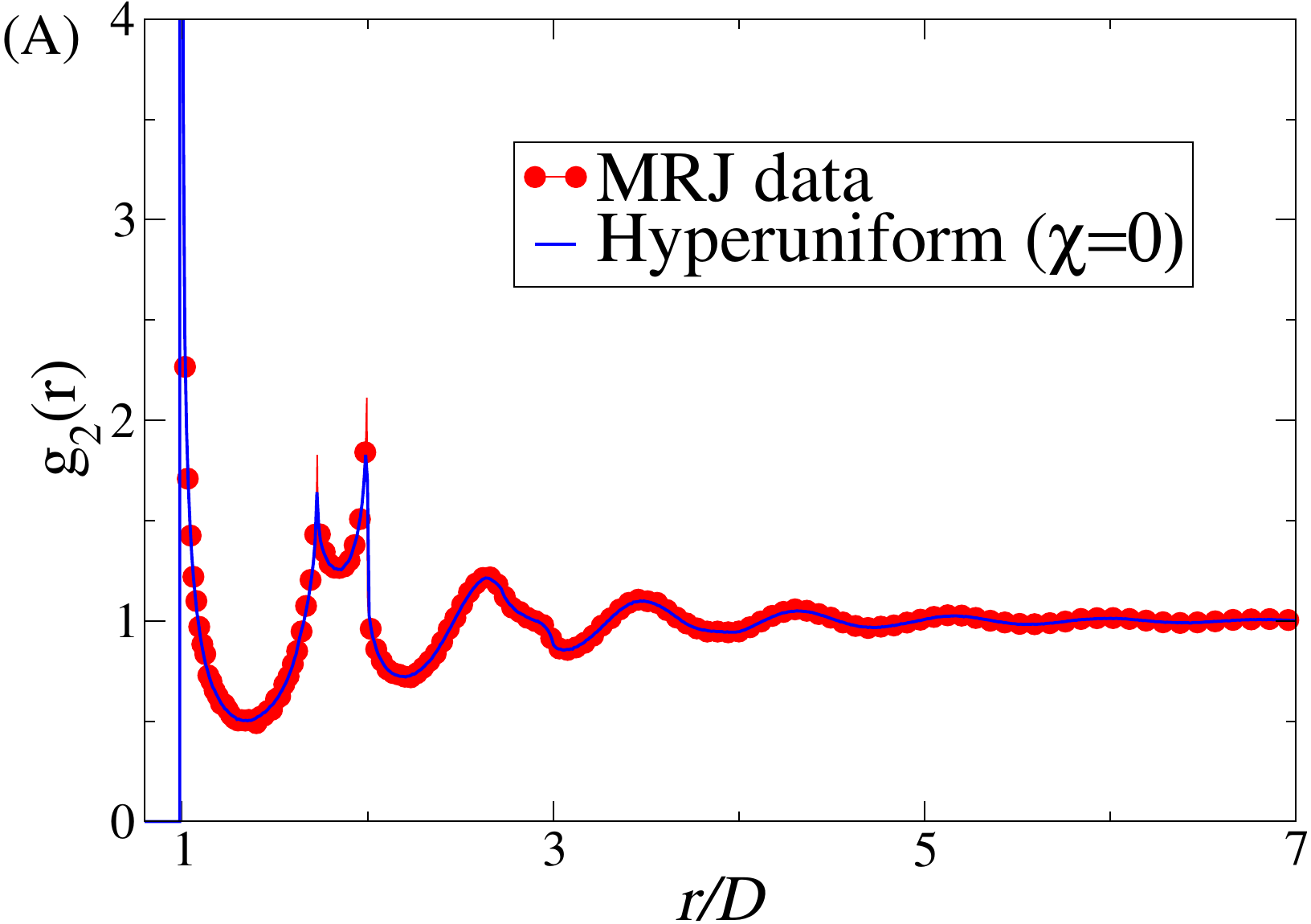}

\includegraphics[width=0.45\textwidth,keepaspectratio,clip=]{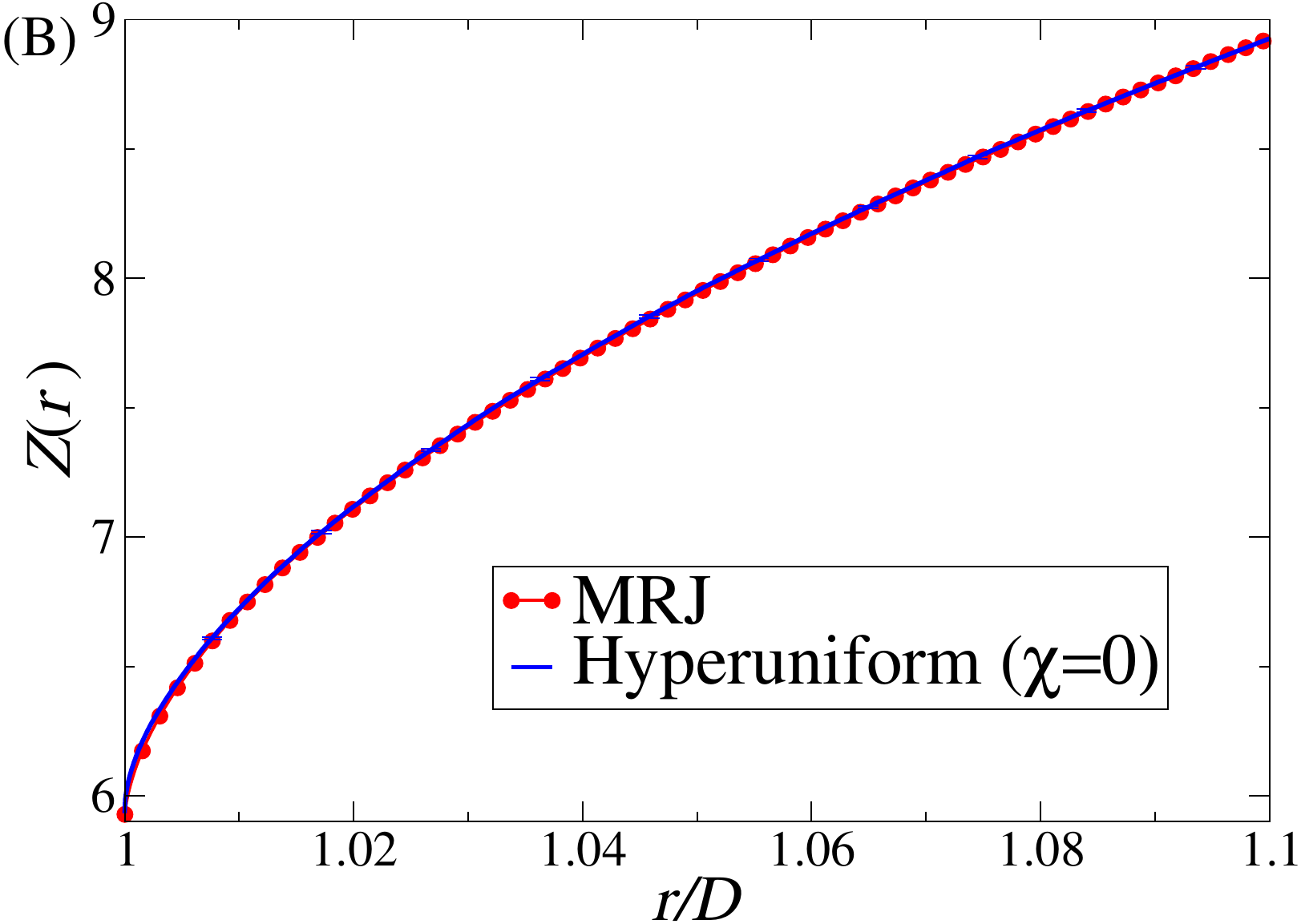}
\caption{Comparisons of 3D hyperuniform sphere-packing ground states ($\chi=0^+$) and 3D MRJ sphere packings in direct space.
(A) The pair correlation function $g_2(r)$ versus dimensionless distance $r/D$ of the two models.
(B) The cumulative coordination number $Z(r)$ versus dimensionless distance $r/D$ of the two models.
In both panels, we consider 3D MRJ packings taken from Ref. \citenum{Ma23}, and $D$ is the hard-sphere diameter.
In both (A) and (B), two models have the same packing fraction, $\phi=0.638$.
}
\label{3D-g2-chi_min}
\end{figure}

To further confirm the striking correspondence to MRJ packings, we compare its
pair correlation function $g_2(r)$ to that of our ground-state packings with $\chi=0^+$ in Fig. \ref{3D-g2-chi_min}(A) for the largest
system size.
Again, we see that the pair correlation functions are virtually identical to one another on the scale of this figure, including
subtle small-scale structural features. Specifically, they share
a Dirac-delta-like peak at $r=D$, the well-known split second peak seen in disordered jammed packings at $r/D=\sqrt{3}$ and $2$, as well as the power-law singularity for near contacts.\cite{Do05c} 
For general dense packings, the exponent $\gamma$ in the power-law singularity, i.e., $g_2(r) \sim (r/D-1)^{-\gamma}$ for $0<r/D-1<0.2$, is expected to increase with increasing short-range order.\cite{Do05c}
Specifically, $\gamma\to0$ means a constant $g_2$ near contact, which is a signature of the ideal gas.
In contrast, $\gamma\to 1$ means that the first and second peaks are clearly separated with a wide range of gaps, which is typical of crystal packings.
Thus, $\gamma$ for the MRJ-like SHU ground states should lie between 0 and 1.

The excellent agreement of the near-contact divergence in the ground-state packings with $\chi=0^+$ and MRJ packings is more apparent in the cumulative coordination number $Z(r)$; see Fig. \ref{3D-g2-chi_min}(B).
Here, we find that both packings
yield the same fit $Z(x) \sim Z_\mathrm{fit} + C x^{1-\gamma}$, where $x\equiv r/D-1$.
In the near-contact region (i.e., $0.001<x<0.1$), the fitted gap exponents are $\gamma=0.44(1)$ for largest ultradense SHU packings with $\chi=\chi_{\min}$ and the MRJ state.
Table \ref{tab:gap} summarizes the values of $\gamma$ at various system sizes $N$.
Interestingly, by the same contact criterion stated in the caption of  Fig. \ref{images}, we estimate the rattler concentration to be about 2\%,
which is close to the recently reported value for MRJ packings.\cite{Ma23}

\begin{figure}
    \centering
    \includegraphics[width=0.45\textwidth]{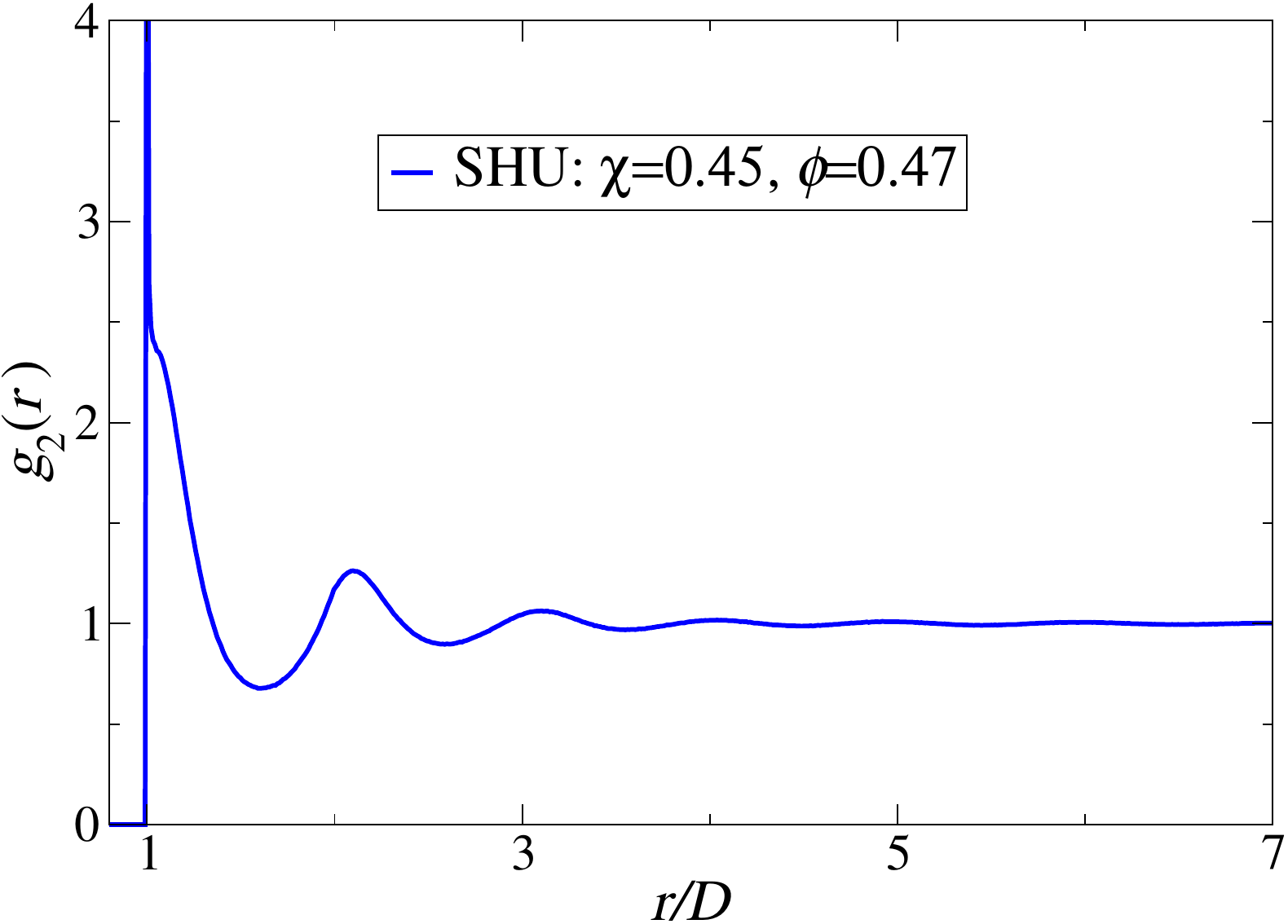}
    \caption{The pair correlation function $g_2(r)$ versus dimensionless distance $r/D$ of 3D SHU sphere-packing ground states with a high value of $\chi$.
    Here, $D$ is the hard-sphere diameter.
    We consider the ground states with $\chi=0.45$, $N=4000$, $\phi=0.47$, and $n_c=500$.
    }
    \label{fig:g2-0.45}
\end{figure}

For the SHU packings with the largest value of the stealthiness parameter examined, i.e., $\chi(=0.45)$, the pair correlation function $g_2(r)$ also possesses a sharp peak at $r=D$ but does not exhibit either a power-law divergence for near contacts nor split second peaks; see Fig. \ref{fig:g2-0.45}.
Furthermore, the small-scale oscillations in $g_2(r)$ for $\chi=0.45$ for large $r$ decay slightly faster than those with $\chi=0^+$.
Such differences in $g_2(r)$ for the SHU ground states with $\chi=0^+$ and $0.45$ arise because those with $\chi=0.45$ form only a small number of dimers but lack additional short-range order.

\section{Conclusions and Discussion}\label{sec:discussion}

In sum, we have shown that one can use the modified collective-coordinate optimization procedure [cf. Eq. \eqref{modified}]
to create disordered SHU ground-state packings with packing fractions that far exceed those produced
in our previous work within the disordered $\chi$ regime,\cite{Ki20a,To21a} thus advancing our understanding of the nature of this highly degenerate ground-state manifold.
We revealed that in the small-$\chi$ limit, the ground-state packings associated with Eq. \eqref{modified} are hyperuniform, effectively jammed, and configurationally very near MRJ sphere packings, as measured by the same values of the packing fraction
and mean contact number per particle, similar rattler concentrations, as well as pair statistics, both $S(k)$ and $g_2(r)$, that are 
strikingly virtually identical to one another. 
The extraordinary existence of nonstealthy hyperuniform MRJ-like packings in the degenerate
SHU ground-state manifold is possible because the cardinality of this infinite set is maximized as the stealthiness parameter w tends to zero.
On the other hand, while our ground-state
packings have a Dirac-delta distribution in packing fraction for finite $N$, numerically created
MRJ packings have a range of packing fractions for finite $N$, which becomes a Dirac-delta distribution only in the thermodynamic limit. 

We also showed that the modified optimization technique yields 
a large family of SHU packings for all values of $\chi<1/2$ that exhibit other novel structural features.
The particles in these disordered packings always form interparticle contacts, albeit with sparser contact networks, and have heretofore unattained maximal packing fractions $\phi_{\max}(\chi)$ compared to previous works on stealthy hyperuniform systems.
In the $\chi$-$\phi_{\max}$ plane, the regions below and above the function $\phi_{\max}(\chi)$ are satisfiable and unsatisfiable phases, respectively; see Fig. \ref{max-phi}(a).
By increasing $\chi$ from $0$ to $1/2$, one can tune the contact networks of the SHU ground states from those that percolate with various degrees of connectivity for small to intermediate $\chi$ values 
to those that do not percolate at higher $\chi$ values characterized by dimer polymer-like chains.
Detailed analyses of local structural statistics of these 3D SHU packings as a function of system size for $\chi >0$ and their lower-dimensional counterparts are reported in Ref. \citenum{Ki25}.

The capability to attain ultradense SHU packings for all $\chi$ within
the disordered regime opens up exploration of new applications in optics and acoustics using such exotic amorphous materials.\cite{Bi19,Roh20,Ki20a,To21a,Zh21,Me23,Al23,Ta23,Ku24}
Indeed, \citealt{Va25} recently showed how certain dynamical physical properties (i.e., effective dynamic dielectric constant and time-dependent diffusion spreadability) of two-phase media derived from them are improved for a range of $\chi$ within the disordered regime and packing fractions $\phi$.

Finally, we note that it has been a great computational challenge to generate the ideal MRJ state, which must be free of rattlers (defects).\cite{To21c}
Rapid-compression algorithms have been used over the years 
to generate such packings,\cite{To00b,Do05d,To10e,Ma23}, but they produce a small but significant fraction of rattlers that prevent the ideal
MRJ state from truly forming due to a critical slowing down. \cite{At16a} In future work, it would be interesting to explore the use of
our effectively jammed but hyperuniform packing states as improved initial conditions for the Torquato-Jiao linear programming packing
algorithm, which is designed to find inherent structures efficiently,\cite{To10e} to possibly
reduce the fraction of rattlers below 1.5\%, thereby approaching the ideal state.

\appendix
\section{Simulation Parameters}
\label{sec:parameters}

Here, we tabulate the simulation parameters for 3D ultradense SHU packings employed in this work.

\begin{table}[h]
\caption{Simulation parameters for 3D ultradense SHU ground-state packings at $\chi$. Here, $N$ is the system size, and $n_c$ is the number of independent ground states.	\label{tab:nc}}
\begin{tabular}{r c r}

	\hline
	$N$&	$\chi$&	$n_c$\\
	\hline
	400  &	$\chi_{\min}=2.51\times10^{-3}$&	10000\\
	1198 &	$\chi_{\min}=8.35\times10^{-4}$&	5000\\
	3592 &	$\chi_{\min}=2.78\times10^{-4}$&	1000\\
	10774 &	$\chi_{\min}=9.28\times10^{-5}$&	100\\
	\hline
    4000    &   0.10    &   500 \\
    4000    &   0.20    &   500 \\
    4000    &   0.30    &   500 \\
    4000    &   0.35    &   500 \\
    4000    &   0.40    &   500 \\
    4000    &   0.45    &   500 \\
    \hline
\end{tabular}
\end{table}

\section{Total Potential in the Limit of $\chi\to 0$}
\label{app:Phi_0}

We provide an explicit expression of the total potential \eqref{modified} with $\chi=\chi_{\min}(N)$ of $N$-particle systems (denoted by $\chi=0^+$ in Sec. \ref{sec:results}):
\begin{align}
    \fn{\Phi}{\vect{r}^N} =&\rho v_0 \left[
		S(k_{\min}\vect{\hat{e}}_1)+S(k_{\min}\vect{\hat{e}}_2)+S(k_{\min}\vect{\hat{e}}_3)
	\right]	
	+ \sum_{i <j} \fn{u}{r_{ij}},
\label{eq:zero-limit}
\end{align}
where $k_{\min} =2\pi/L$ is the smallest wavenumber of the reciprocal lattice vectors for the periodic simulation box, and $\vect{\hat{e}}_i$ ($i=1,2,3$) are the standard basis of $\R^3$.
Importantly, we approach the limit of $\chi\to0$ by increasing $N$ in Eq. \eqref{eq:zero-limit} as large as possible (i.e., $N=10774$), ensuring that the obtained ground states are hyperuniform in this limit, which is challenging for conventional compression algorithms to achieve.

\section*{Author contributions}
S.T.: conceptualization, investigation, data curation, writing, funding acquisition; J.K.: data curation, software, investigation, writing.

\section*{Conflicts of interest}
There are no conflicts to declare.

\section*{Data availability}
The data that support the findings of this study are available from the corresponding author upon reasonable request.

\section*{Acknowledgements}
We thank Peter Morse, Charles Maher, and Paul Steinhardt for their very helpful remarks.
This research was sponsored by the Army Research Office, accomplished under Cooperative Agreement Number W911NF-22-2-0103 and the National Science Foundation under Award No. CBET-2133179.






\begin{mcitethebibliography}{80}
\providecommand*{\natexlab}[1]{#1}
\providecommand*{\mciteSetBstSublistMode}[1]{}
\providecommand*{\mciteSetBstMaxWidthForm}[2]{}
\providecommand*{\mciteBstWouldAddEndPuncttrue}
  {\def\EndOfBibitem{\unskip.}}
\providecommand*{\mciteBstWouldAddEndPunctfalse}
  {\let\EndOfBibitem\relax}
\providecommand*{\mciteSetBstMidEndSepPunct}[3]{}
\providecommand*{\mciteSetBstSublistLabelBeginEnd}[3]{}
\providecommand*{\EndOfBibitem}{}
\mciteSetBstSublistMode{f}
\mciteSetBstMaxWidthForm{subitem}
{(\emph{\alph{mcitesubitemcount}})}
\mciteSetBstSublistLabelBeginEnd{\mcitemaxwidthsubitemform\space}
{\relax}{\relax}

\bibitem[Torquato and Stillinger(2003)]{To03a}
S.~Torquato and F.~H. Stillinger, \emph{Phys. Rev. E}, 2003, \textbf{68},
  041113\relax
\mciteBstWouldAddEndPuncttrue
\mciteSetBstMidEndSepPunct{\mcitedefaultmidpunct}
{\mcitedefaultendpunct}{\mcitedefaultseppunct}\relax
\EndOfBibitem
\bibitem[Torquato(2018)]{To18a}
S.~Torquato, \emph{Phys. Rep.}, 2018, \textbf{745}, 1--95\relax
\mciteBstWouldAddEndPuncttrue
\mciteSetBstMidEndSepPunct{\mcitedefaultmidpunct}
{\mcitedefaultendpunct}{\mcitedefaultseppunct}\relax
\EndOfBibitem
\bibitem[Dyson(1970)]{Dy70}
F.~J. Dyson, \emph{Comm. Math. Phys.}, 1970, \textbf{19}, 235--250\relax
\mciteBstWouldAddEndPuncttrue
\mciteSetBstMidEndSepPunct{\mcitedefaultmidpunct}
{\mcitedefaultendpunct}{\mcitedefaultseppunct}\relax
\EndOfBibitem
\bibitem[Montgomery(1973)]{Mon73}
H.~L. Montgomery, \emph{Amer. Math. Soc.}, 1973,  181--193\relax
\mciteBstWouldAddEndPuncttrue
\mciteSetBstMidEndSepPunct{\mcitedefaultmidpunct}
{\mcitedefaultendpunct}{\mcitedefaultseppunct}\relax
\EndOfBibitem
\bibitem[Donev \emph{et~al.}(2005)Donev, Stillinger, and Torquato]{Do05d}
A.~Donev, F.~H. Stillinger and S.~Torquato, \emph{Phys. Rev. Lett.}, 2005,
  \textbf{95}, 090604\relax
\mciteBstWouldAddEndPuncttrue
\mciteSetBstMidEndSepPunct{\mcitedefaultmidpunct}
{\mcitedefaultendpunct}{\mcitedefaultseppunct}\relax
\EndOfBibitem
\bibitem[Maher \emph{et~al.}(2023)Maher, Jiao, and Torquato]{Ma23}
C.~E. Maher, Y.~Jiao and S.~Torquato, \emph{Phys. Rev. E}, 2023, \textbf{108},
  064602\relax
\mciteBstWouldAddEndPuncttrue
\mciteSetBstMidEndSepPunct{\mcitedefaultmidpunct}
{\mcitedefaultendpunct}{\mcitedefaultseppunct}\relax
\EndOfBibitem
\bibitem[Jiao \emph{et~al.}(2014)Jiao, Lau, Hatzikirou, Meyer-Hermann, Corbo,
  and Torquato]{Ji14}
Y.~Jiao, T.~Lau, H.~Hatzikirou, M.~Meyer-Hermann, J.~C. Corbo and S.~Torquato,
  \emph{Phys. Rev. E}, 2014, \textbf{89}, 022721\relax
\mciteBstWouldAddEndPuncttrue
\mciteSetBstMidEndSepPunct{\mcitedefaultmidpunct}
{\mcitedefaultendpunct}{\mcitedefaultseppunct}\relax
\EndOfBibitem
\bibitem[Mayer \emph{et~al.}(2015)Mayer, Balasubramanian, Mora, and
  Walczak]{Ma15}
A.~Mayer, V.~Balasubramanian, T.~Mora and A.~M. Walczak, \emph{Proc. Nat. Acad.
  Sci. U. S. A.}, 2015, \textbf{112}, 5950--5955\relax
\mciteBstWouldAddEndPuncttrue
\mciteSetBstMidEndSepPunct{\mcitedefaultmidpunct}
{\mcitedefaultendpunct}{\mcitedefaultseppunct}\relax
\EndOfBibitem
\bibitem[Torquato \emph{et~al.}(2008)Torquato, Scardicchio, and Zachary]{To08b}
S.~Torquato, A.~Scardicchio and C.~E. Zachary, \emph{J. Stat. Mech.: Theory
  Exp.}, 2008, \textbf{2008}, P11019\relax
\mciteBstWouldAddEndPuncttrue
\mciteSetBstMidEndSepPunct{\mcitedefaultmidpunct}
{\mcitedefaultendpunct}{\mcitedefaultseppunct}\relax
\EndOfBibitem
\bibitem[Wang \emph{et~al.}(2024)Wang, Samajdar, and Torquato]{Wa24c}
H.~Wang, R.~Samajdar and S.~Torquato, \emph{Phys. Rev. B}, 2024, \textbf{110},
  104201\relax
\mciteBstWouldAddEndPuncttrue
\mciteSetBstMidEndSepPunct{\mcitedefaultmidpunct}
{\mcitedefaultendpunct}{\mcitedefaultseppunct}\relax
\EndOfBibitem
\bibitem[Reichhardt and Reichhardt(2014)]{Re14}
C.~Reichhardt and C.~J.~O. Reichhardt, \emph{Soft Matter}, 2014, \textbf{10},
  7502--7510\relax
\mciteBstWouldAddEndPuncttrue
\mciteSetBstMidEndSepPunct{\mcitedefaultmidpunct}
{\mcitedefaultendpunct}{\mcitedefaultseppunct}\relax
\EndOfBibitem
\bibitem[{Hexner} and {Levine}(2015)]{He15}
D.~{Hexner} and D.~{Levine}, \emph{Phys. Rev. Lett.}, 2015, \textbf{114},
  110602\relax
\mciteBstWouldAddEndPuncttrue
\mciteSetBstMidEndSepPunct{\mcitedefaultmidpunct}
{\mcitedefaultendpunct}{\mcitedefaultseppunct}\relax
\EndOfBibitem
\bibitem[Hexner \emph{et~al.}(2017)Hexner, Chaikin, and Levine]{He17b}
D.~Hexner, P.~M. Chaikin and D.~Levine, \emph{Proc. Nat. Acad. Sci. U. S. A.},
  2017, \textbf{114}, 4294--4299\relax
\mciteBstWouldAddEndPuncttrue
\mciteSetBstMidEndSepPunct{\mcitedefaultmidpunct}
{\mcitedefaultendpunct}{\mcitedefaultseppunct}\relax
\EndOfBibitem
\bibitem[Zhang \emph{et~al.}(2023)Zhang, Wang, Zhang, Yu, and Douglas]{Zh23}
H.~Zhang, X.~Wang, J.~Zhang, H.-B. Yu and J.~F. Douglas, \emph{Eur. Phys. J.
  E}, 2023, \textbf{46}, 50\relax
\mciteBstWouldAddEndPuncttrue
\mciteSetBstMidEndSepPunct{\mcitedefaultmidpunct}
{\mcitedefaultendpunct}{\mcitedefaultseppunct}\relax
\EndOfBibitem
\bibitem[Lei and Ni(2019)]{Le19a}
Q.-L. Lei and R.~Ni, \emph{Proc. Nat. Acad. Sci.}, 2019, \textbf{116},
  22983--22989\relax
\mciteBstWouldAddEndPuncttrue
\mciteSetBstMidEndSepPunct{\mcitedefaultmidpunct}
{\mcitedefaultendpunct}{\mcitedefaultseppunct}\relax
\EndOfBibitem
\bibitem[Zachary and Torquato(2009)]{Za09}
C.~E. Zachary and S.~Torquato, \emph{J. Stat. Mech.: Theory \& Exp.}, 2009,
  \textbf{2009}, P12015\relax
\mciteBstWouldAddEndPuncttrue
\mciteSetBstMidEndSepPunct{\mcitedefaultmidpunct}
{\mcitedefaultendpunct}{\mcitedefaultseppunct}\relax
\EndOfBibitem
\bibitem[{O{\u g}uz} \emph{et~al.}(2017){O{\u g}uz}, {Socolar}, {Steinhardt},
  and {Torquato}]{Og17}
E.~C. {O{\u g}uz}, J.~E.~S. {Socolar}, P.~J. {Steinhardt} and S.~{Torquato},
  \emph{Phys. Rev. B}, 2017, \textbf{95}, 054119\relax
\mciteBstWouldAddEndPuncttrue
\mciteSetBstMidEndSepPunct{\mcitedefaultmidpunct}
{\mcitedefaultendpunct}{\mcitedefaultseppunct}\relax
\EndOfBibitem
\bibitem[Torquato \emph{et~al.}(2019)Torquato, Zhang, and
  de~Courcy-Ireland]{To19}
S.~Torquato, G.~Zhang and M.~de~Courcy-Ireland, \emph{J. Phys. A: Math. \&
  Theoretical}, 2019, \textbf{52}, 135002\relax
\mciteBstWouldAddEndPuncttrue
\mciteSetBstMidEndSepPunct{\mcitedefaultmidpunct}
{\mcitedefaultendpunct}{\mcitedefaultseppunct}\relax
\EndOfBibitem
\bibitem[Philcox and Torquato(2023)]{Ph23}
O.~H.~E. Philcox and S.~Torquato, \emph{Phys. Rev. X}, 2023, \textbf{13},
  011038\relax
\mciteBstWouldAddEndPuncttrue
\mciteSetBstMidEndSepPunct{\mcitedefaultmidpunct}
{\mcitedefaultendpunct}{\mcitedefaultseppunct}\relax
\EndOfBibitem
\bibitem[Chremos and Douglas(2018)]{Ch18}
A.~Chremos and J.~F. Douglas, \emph{Phys. Rev. Lett.}, 2018, \textbf{121},
  258002\relax
\mciteBstWouldAddEndPuncttrue
\mciteSetBstMidEndSepPunct{\mcitedefaultmidpunct}
{\mcitedefaultendpunct}{\mcitedefaultseppunct}\relax
\EndOfBibitem
\bibitem[Chen \emph{et~al.}(2025)Chen, Samajdar, Jiao, and Torquato]{Ch25}
D.~Chen, R.~Samajdar, Y.~Jiao and S.~Torquato, \emph{Proc. Nat. Acad. Soc.},
  2025, \textbf{122}, e2416111122\relax
\mciteBstWouldAddEndPuncttrue
\mciteSetBstMidEndSepPunct{\mcitedefaultmidpunct}
{\mcitedefaultendpunct}{\mcitedefaultseppunct}\relax
\EndOfBibitem
\bibitem[Uche \emph{et~al.}(2004)Uche, Stillinger, and Torquato]{Uc04b}
O.~U. Uche, F.~H. Stillinger and S.~Torquato, \emph{Phys. Rev. E}, 2004,
  \textbf{70}, 046122\relax
\mciteBstWouldAddEndPuncttrue
\mciteSetBstMidEndSepPunct{\mcitedefaultmidpunct}
{\mcitedefaultendpunct}{\mcitedefaultseppunct}\relax
\EndOfBibitem
\bibitem[Batten \emph{et~al.}(2008)Batten, Stillinger, and Torquato]{Ba08}
R.~D. Batten, F.~H. Stillinger and S.~Torquato, \emph{J. Appl. Phys.}, 2008,
  \textbf{104}, 033504\relax
\mciteBstWouldAddEndPuncttrue
\mciteSetBstMidEndSepPunct{\mcitedefaultmidpunct}
{\mcitedefaultendpunct}{\mcitedefaultseppunct}\relax
\EndOfBibitem
\bibitem[{Torquato} \emph{et~al.}(2015){Torquato}, {Zhang}, and
  {Stillinger}]{To15}
S.~{Torquato}, G.~{Zhang} and F.~H. {Stillinger}, \emph{Phys. Rev. X}, 2015,
  \textbf{5}, 021020\relax
\mciteBstWouldAddEndPuncttrue
\mciteSetBstMidEndSepPunct{\mcitedefaultmidpunct}
{\mcitedefaultendpunct}{\mcitedefaultseppunct}\relax
\EndOfBibitem
\bibitem[Zhang \emph{et~al.}(2017)Zhang, Stillinger, and Torquato]{Zh17a}
G.~Zhang, F.~H. Stillinger and S.~Torquato, \emph{Soft Matter}, 2017,
  \textbf{13}, 6197--6207\relax
\mciteBstWouldAddEndPuncttrue
\mciteSetBstMidEndSepPunct{\mcitedefaultmidpunct}
{\mcitedefaultendpunct}{\mcitedefaultseppunct}\relax
\EndOfBibitem
\bibitem[{Ghosh} and {Lebowitz}(2018)]{Gh18}
S.~{Ghosh} and J.~L. {Lebowitz}, \emph{Comm. Math. Phys.}, 2018, \textbf{363},
  97--110\relax
\mciteBstWouldAddEndPuncttrue
\mciteSetBstMidEndSepPunct{\mcitedefaultmidpunct}
{\mcitedefaultendpunct}{\mcitedefaultseppunct}\relax
\EndOfBibitem
\bibitem[Salvalaglio \emph{et~al.}(2024)Salvalaglio, Skinner, Dunkel, and
  Voigt]{Sa24}
M.~Salvalaglio, D.~J. Skinner, J.~Dunkel and A.~Voigt, \emph{Phys. Rev. Res.},
  2024, \textbf{6}, 023107\relax
\mciteBstWouldAddEndPuncttrue
\mciteSetBstMidEndSepPunct{\mcitedefaultmidpunct}
{\mcitedefaultendpunct}{\mcitedefaultseppunct}\relax
\EndOfBibitem
\bibitem[Florescu \emph{et~al.}(2009)Florescu, Torquato, and Steinhardt]{Fl09b}
M.~Florescu, S.~Torquato and P.~J. Steinhardt, \emph{Proc. Nat. Acad. Sci. U.
  S. A.}, 2009, \textbf{106}, 20658--20663\relax
\mciteBstWouldAddEndPuncttrue
\mciteSetBstMidEndSepPunct{\mcitedefaultmidpunct}
{\mcitedefaultendpunct}{\mcitedefaultseppunct}\relax
\EndOfBibitem
\bibitem[Gkantzounis \emph{et~al.}(2017)Gkantzounis, Amoah, and Florescu]{Gk17}
G.~Gkantzounis, T.~Amoah and M.~Florescu, \emph{Phys. Rev. B}, 2017,
  \textbf{95}, 094120\relax
\mciteBstWouldAddEndPuncttrue
\mciteSetBstMidEndSepPunct{\mcitedefaultmidpunct}
{\mcitedefaultendpunct}{\mcitedefaultseppunct}\relax
\EndOfBibitem
\bibitem[{Froufe-P{\'e}rez} \emph{et~al.}(2017){Froufe-P{\'e}rez}, {Engel},
  {Jos{\'e} S{\'a}enz}, and {Scheffold}]{Fr17}
L.~S. {Froufe-P{\'e}rez}, M.~{Engel}, J.~{Jos{\'e} S{\'a}enz} and
  F.~{Scheffold}, \emph{Proc. Nat. Acad. Sci.}, 2017, \textbf{114},
  9570--9574\relax
\mciteBstWouldAddEndPuncttrue
\mciteSetBstMidEndSepPunct{\mcitedefaultmidpunct}
{\mcitedefaultendpunct}{\mcitedefaultseppunct}\relax
\EndOfBibitem
\bibitem[Castro-Lopez \emph{et~al.}(2017)Castro-Lopez, Gaio, Sellers,
  Gkantzounis, Florescu, and Sapienza]{Cas17}
M.~Castro-Lopez, M.~Gaio, S.~Sellers, G.~Gkantzounis, M.~Florescu and
  R.~Sapienza, \emph{APL Photonics}, 2017, \textbf{2}, 061302\relax
\mciteBstWouldAddEndPuncttrue
\mciteSetBstMidEndSepPunct{\mcitedefaultmidpunct}
{\mcitedefaultendpunct}{\mcitedefaultseppunct}\relax
\EndOfBibitem
\bibitem[Torquato and Chen(2018)]{To18c}
S.~Torquato and D.~Chen, \emph{Multifunctional Materials}, 2018, \textbf{1},
  015001\relax
\mciteBstWouldAddEndPuncttrue
\mciteSetBstMidEndSepPunct{\mcitedefaultmidpunct}
{\mcitedefaultendpunct}{\mcitedefaultseppunct}\relax
\EndOfBibitem
\bibitem[Bigourdan \emph{et~al.}(2019)Bigourdan, Pierrat, and Carminati]{Bi19}
F.~Bigourdan, R.~Pierrat and R.~Carminati, \emph{Optics Express}, 2019,
  \textbf{27}, 8666--8682\relax
\mciteBstWouldAddEndPuncttrue
\mciteSetBstMidEndSepPunct{\mcitedefaultmidpunct}
{\mcitedefaultendpunct}{\mcitedefaultseppunct}\relax
\EndOfBibitem
\bibitem[{Romero-Garc{\'i}a} \emph{et~al.}(2019){Romero-Garc{\'i}a}, Lamothe,
  Theocharis, Richoux, and {Garc{\'i}a-Raffi}]{Ro19}
V.~{Romero-Garc{\'i}a}, N.~Lamothe, G.~Theocharis, O.~Richoux and L.~M.
  {Garc{\'i}a-Raffi}, \emph{Phys. Rev. Appl.}, 2019, \textbf{11}, 054076\relax
\mciteBstWouldAddEndPuncttrue
\mciteSetBstMidEndSepPunct{\mcitedefaultmidpunct}
{\mcitedefaultendpunct}{\mcitedefaultseppunct}\relax
\EndOfBibitem
\bibitem[Rohfritsch \emph{et~al.}(2020)Rohfritsch, Conoir, {Valier-Brasier},
  and Marchiano]{Roh20}
A.~Rohfritsch, J.-M. Conoir, T.~{Valier-Brasier} and R.~Marchiano, \emph{Phys.
  Rev. E}, 2020, \textbf{102}, 053001\relax
\mciteBstWouldAddEndPuncttrue
\mciteSetBstMidEndSepPunct{\mcitedefaultmidpunct}
{\mcitedefaultendpunct}{\mcitedefaultseppunct}\relax
\EndOfBibitem
\bibitem[Torquato and Kim(2021)]{To21a}
S.~Torquato and J.~Kim, \emph{Phys. Rev. X}, 2021, \textbf{11}, 021002\relax
\mciteBstWouldAddEndPuncttrue
\mciteSetBstMidEndSepPunct{\mcitedefaultmidpunct}
{\mcitedefaultendpunct}{\mcitedefaultseppunct}\relax
\EndOfBibitem
\bibitem[Christogeorgos \emph{et~al.}(2021)Christogeorgos, Zhang, Cheng, and
  Hao]{Ch21}
O.~Christogeorgos, H.~Zhang, Q.~Cheng and Y.~Hao, \emph{Phys. Rev. Appl.},
  2021, \textbf{15}, 014062\relax
\mciteBstWouldAddEndPuncttrue
\mciteSetBstMidEndSepPunct{\mcitedefaultmidpunct}
{\mcitedefaultendpunct}{\mcitedefaultseppunct}\relax
\EndOfBibitem
\bibitem[Zhang \emph{et~al.}(2021)Zhang, Cheng, Chu, Christogeorgos, Wu, and
  Hao]{Zh21}
H.~Zhang, Q.~Cheng, H.~Chu, O.~Christogeorgos, W.~Wu and Y.~Hao, \emph{Appl.
  Phys. Lett.}, 2021, \textbf{118}, 101601\relax
\mciteBstWouldAddEndPuncttrue
\mciteSetBstMidEndSepPunct{\mcitedefaultmidpunct}
{\mcitedefaultendpunct}{\mcitedefaultseppunct}\relax
\EndOfBibitem
\bibitem[{Romero-Garc{\'i}a} \emph{et~al.}(2021){Romero-Garc{\'i}a},
  Ch{\'e}ron, Kuznetsova, Groby, F{\'e}lix, Pagneux, and {Garcia-Raffi}]{Ro21}
V.~{Romero-Garc{\'i}a}, {\'E}.~Ch{\'e}ron, S.~Kuznetsova, J.-P. Groby,
  S.~F{\'e}lix, V.~Pagneux and L.~M. {Garcia-Raffi}, \emph{APL Mater.}, 2021,
  \textbf{9}, 101101\relax
\mciteBstWouldAddEndPuncttrue
\mciteSetBstMidEndSepPunct{\mcitedefaultmidpunct}
{\mcitedefaultendpunct}{\mcitedefaultseppunct}\relax
\EndOfBibitem
\bibitem[Klatt \emph{et~al.}(2022)Klatt, Steinhardt, and Torquato]{Kl22}
M.~A. Klatt, P.~J. Steinhardt and S.~Torquato, \emph{Proc. Nat. Acad. Sci.},
  2022, \textbf{119}, e2213633119\relax
\mciteBstWouldAddEndPuncttrue
\mciteSetBstMidEndSepPunct{\mcitedefaultmidpunct}
{\mcitedefaultendpunct}{\mcitedefaultseppunct}\relax
\EndOfBibitem
\bibitem[Tavakoli \emph{et~al.}(2022)Tavakoli, Spalding, Lambertz, Koppejan,
  Gkantzounis, Wan, R{"o}hrich, Kontoleta, Koenderink, Sapienza, Florescu, and
  Alarcon-Llado]{Ta22}
N.~Tavakoli, R.~Spalding, A.~Lambertz, P.~Koppejan, G.~Gkantzounis, C.~Wan,
  R.~R{"o}hrich, E.~Kontoleta, A.~F. Koenderink, R.~Sapienza, M.~Florescu and
  E.~Alarcon-Llado, \emph{ACS photonics}, 2022, \textbf{9}, 1206--1217\relax
\mciteBstWouldAddEndPuncttrue
\mciteSetBstMidEndSepPunct{\mcitedefaultmidpunct}
{\mcitedefaultendpunct}{\mcitedefaultseppunct}\relax
\EndOfBibitem
\bibitem[Tang \emph{et~al.}(2023)Tang, Wang, Wang, Gao, Li, Liang, Sebbah, Li,
  Zhang, and Shi]{Ta23}
K.~Tang, Y.~Wang, S.~Wang, D.~Gao, H.~Li, X.~Liang, P.~Sebbah, Y.~Li, J.~Zhang
  and J.~Shi, \emph{Phys. Rev. Appl.}, 2023, \textbf{19}, 054035\relax
\mciteBstWouldAddEndPuncttrue
\mciteSetBstMidEndSepPunct{\mcitedefaultmidpunct}
{\mcitedefaultendpunct}{\mcitedefaultseppunct}\relax
\EndOfBibitem
\bibitem[Merkel \emph{et~al.}(2023)Merkel, Stappers, Ray, Denz, and
  Imbrock]{Me23}
M.~Merkel, M.~Stappers, D.~Ray, C.~Denz and J.~Imbrock, \emph{Adv. Photonics
  Res.}, 2023, \textbf{5}, 2300256\relax
\mciteBstWouldAddEndPuncttrue
\mciteSetBstMidEndSepPunct{\mcitedefaultmidpunct}
{\mcitedefaultendpunct}{\mcitedefaultseppunct}\relax
\EndOfBibitem
\bibitem[Alha{\"\i}tz \emph{et~al.}(2023)Alha{\"\i}tz, Conoir, and
  Valier-Brasier]{Al23}
L.~Alha{\"\i}tz, J.-M. Conoir and T.~Valier-Brasier, \emph{Phys. Rev. E}, 2023,
  \textbf{108}, 065001\relax
\mciteBstWouldAddEndPuncttrue
\mciteSetBstMidEndSepPunct{\mcitedefaultmidpunct}
{\mcitedefaultendpunct}{\mcitedefaultseppunct}\relax
\EndOfBibitem
\bibitem[Granchi \emph{et~al.}(2023)Granchi, Lodde, Stokkereit, Spalding,
  Veldhoven, Sapienza, Fiore, Gurioli, Florescu, and Intonti]{Gr23b}
N.~Granchi, M.~Lodde, K.~Stokkereit, R.~Spalding, P.~J. Veldhoven, R.~Sapienza,
  A.~Fiore, M.~Gurioli, M.~Florescu and F.~Intonti, \emph{Phys. Rev. B}, 2023,
  \textbf{107}, 064204\relax
\mciteBstWouldAddEndPuncttrue
\mciteSetBstMidEndSepPunct{\mcitedefaultmidpunct}
{\mcitedefaultendpunct}{\mcitedefaultseppunct}\relax
\EndOfBibitem
\bibitem[Zhuang \emph{et~al.}(2024)Zhuang, Chen, Liu, Keeney, Zhang, and
  Jiao]{Zh24}
H.~Zhuang, D.~Chen, L.~Liu, D.~Keeney, G.~Zhang and Y.~Jiao, \emph{J. Phys.:
  Condens. Matter}, 2024, \textbf{36}, 285703\relax
\mciteBstWouldAddEndPuncttrue
\mciteSetBstMidEndSepPunct{\mcitedefaultmidpunct}
{\mcitedefaultendpunct}{\mcitedefaultseppunct}\relax
\EndOfBibitem
\bibitem[Kuznetsova \emph{et~al.}(2024)Kuznetsova, Groby, Garcia-Raffi, and
  Romero-Garc\'{i}a]{Ku24}
S.~Kuznetsova, J.~P. Groby, L.~M. Garcia-Raffi and V.~Romero-Garc\'{i}a,
  \emph{Waves Random Complex Media}, 2024, \textbf{34}, 1878--1896\relax
\mciteBstWouldAddEndPuncttrue
\mciteSetBstMidEndSepPunct{\mcitedefaultmidpunct}
{\mcitedefaultendpunct}{\mcitedefaultseppunct}\relax
\EndOfBibitem
\bibitem[Wigner(1934)]{Wi36}
E.~Wigner, \emph{Phys. Rev.}, 1934, \textbf{46}, 1002\relax
\mciteBstWouldAddEndPuncttrue
\mciteSetBstMidEndSepPunct{\mcitedefaultmidpunct}
{\mcitedefaultendpunct}{\mcitedefaultseppunct}\relax
\EndOfBibitem
\bibitem[Feynman and Cohen(1956)]{Fe56}
R.~P. Feynman and M.~Cohen, \emph{Phys. Rev.}, 1956, \textbf{102},
  1189--1204\relax
\mciteBstWouldAddEndPuncttrue
\mciteSetBstMidEndSepPunct{\mcitedefaultmidpunct}
{\mcitedefaultendpunct}{\mcitedefaultseppunct}\relax
\EndOfBibitem
\bibitem[Laughlin(1987)]{Laugh87}
R.~B. Laughlin, \emph{The Quantum Hall Effect}, Springer, 1987, pp.
  233--301\relax
\mciteBstWouldAddEndPuncttrue
\mciteSetBstMidEndSepPunct{\mcitedefaultmidpunct}
{\mcitedefaultendpunct}{\mcitedefaultseppunct}\relax
\EndOfBibitem
\bibitem[Broholm \emph{et~al.}(2020)Broholm, Cava, Kivelson, Nocera, Norman,
  and Senthil]{Br20}
C.~Broholm, R.~J. Cava, S.~A. Kivelson, D.~G. Nocera, M.~R. Norman and
  T.~Senthil, \emph{Science}, 2020, \textbf{367}, eaay0668\relax
\mciteBstWouldAddEndPuncttrue
\mciteSetBstMidEndSepPunct{\mcitedefaultmidpunct}
{\mcitedefaultendpunct}{\mcitedefaultseppunct}\relax
\EndOfBibitem
\bibitem[Semeghini \emph{et~al.}(2021)Semeghini, Levine, Keesling, Ebadi, Wang,
  Bluvstein, Verresen, Pichler, Kalinowski, Samajdar, Omran, Sachdev,
  Vishwanath, Greiner, Vuletić, and Lukin]{Se21}
G.~Semeghini, H.~Levine, A.~Keesling, S.~Ebadi, T.~T. Wang, D.~Bluvstein,
  R.~Verresen, H.~Pichler, M.~Kalinowski, R.~Samajdar, A.~Omran, S.~Sachdev,
  A.~Vishwanath, M.~Greiner, V.~Vuletić and M.~D. Lukin, \emph{Science}, 2021,
  \textbf{374}, 1242--1247\relax
\mciteBstWouldAddEndPuncttrue
\mciteSetBstMidEndSepPunct{\mcitedefaultmidpunct}
{\mcitedefaultendpunct}{\mcitedefaultseppunct}\relax
\EndOfBibitem
\bibitem[Samajdar \emph{et~al.}(2021)Samajdar, Ho, Pichler, Lukin, and
  Sachdev]{Sa21}
R.~Samajdar, W.~W. Ho, H.~Pichler, M.~D. Lukin and S.~Sachdev, \emph{Proc. Nat.
  Acad. Soc.}, 2021, \textbf{118}, e2015785118\relax
\mciteBstWouldAddEndPuncttrue
\mciteSetBstMidEndSepPunct{\mcitedefaultmidpunct}
{\mcitedefaultendpunct}{\mcitedefaultseppunct}\relax
\EndOfBibitem
\bibitem[Kivelson and Sondhi(2023)]{Kiv23}
S.~Kivelson and S.~Sondhi, \emph{Nat Rev Phys}, 2023, \textbf{5},
  368--369\relax
\mciteBstWouldAddEndPuncttrue
\mciteSetBstMidEndSepPunct{\mcitedefaultmidpunct}
{\mcitedefaultendpunct}{\mcitedefaultseppunct}\relax
\EndOfBibitem
\bibitem[Zhang \emph{et~al.}(2015)Zhang, Stillinger, and Torquato]{Zh15a}
G.~Zhang, F.~Stillinger and S.~Torquato, \emph{Phys. Rev. E}, 2015,
  \textbf{92}, 022119\relax
\mciteBstWouldAddEndPuncttrue
\mciteSetBstMidEndSepPunct{\mcitedefaultmidpunct}
{\mcitedefaultendpunct}{\mcitedefaultseppunct}\relax
\EndOfBibitem
\bibitem[Morse \emph{et~al.}(2023)Morse, Kim, Steinhardt, and Torquato]{Mo23}
P.~K. Morse, J.~Kim, P.~J. Steinhardt and S.~Torquato, \emph{Phys. Rev. Res.},
  2023, \textbf{5}, 033190\relax
\mciteBstWouldAddEndPuncttrue
\mciteSetBstMidEndSepPunct{\mcitedefaultmidpunct}
{\mcitedefaultendpunct}{\mcitedefaultseppunct}\relax
\EndOfBibitem
\bibitem[Shih \emph{et~al.}(2024)Shih, Casiulis, and Martiniani]{Sh24}
A.~Shih, M.~Casiulis and S.~Martiniani, \emph{Phys. Rev. E}, 2024,
  \textbf{110}, 034122\relax
\mciteBstWouldAddEndPuncttrue
\mciteSetBstMidEndSepPunct{\mcitedefaultmidpunct}
{\mcitedefaultendpunct}{\mcitedefaultseppunct}\relax
\EndOfBibitem
\bibitem[Zhang \emph{et~al.}(2016)Zhang, Stillinger, and Torquato]{Zh16b}
G.~Zhang, F.~H. Stillinger and S.~Torquato, \emph{J. Chem. Phys}, 2016,
  \textbf{145}, 244109\relax
\mciteBstWouldAddEndPuncttrue
\mciteSetBstMidEndSepPunct{\mcitedefaultmidpunct}
{\mcitedefaultendpunct}{\mcitedefaultseppunct}\relax
\EndOfBibitem
\bibitem[Kim and Torquato(2025)]{Ki25}
J.~Kim and S.~Torquato, \emph{J. Chem. Phys., in press; arXiv preprint
  arXiv:2504.16924}, 2025\relax
\mciteBstWouldAddEndPuncttrue
\mciteSetBstMidEndSepPunct{\mcitedefaultmidpunct}
{\mcitedefaultendpunct}{\mcitedefaultseppunct}\relax
\EndOfBibitem
\bibitem[Torquato and Jiao(2010)]{To10e}
S.~Torquato and Y.~Jiao, \emph{Phys. Rev. E}, 2010, \textbf{82}, 061302\relax
\mciteBstWouldAddEndPuncttrue
\mciteSetBstMidEndSepPunct{\mcitedefaultmidpunct}
{\mcitedefaultendpunct}{\mcitedefaultseppunct}\relax
\EndOfBibitem
\bibitem[Charbonneau \emph{et~al.}(2012)Charbonneau, Corwin, Parisi, and
  Zamponi]{Ch12}
P.~Charbonneau, E.~I. Corwin, G.~Parisi and F.~Zamponi, \emph{Phys. Rev.
  Lett.}, 2012, \textbf{109}, 205501\relax
\mciteBstWouldAddEndPuncttrue
\mciteSetBstMidEndSepPunct{\mcitedefaultmidpunct}
{\mcitedefaultendpunct}{\mcitedefaultseppunct}\relax
\EndOfBibitem
\bibitem[Stillinger \emph{et~al.}(1964)Stillinger, Di{M}arzio, and
  Kornegay]{St64b}
F.~H. Stillinger, E.~A. Di{M}arzio and R.~L. Kornegay, \emph{J. Chem. Phys.},
  1964, \textbf{40}, 1564--576\relax
\mciteBstWouldAddEndPuncttrue
\mciteSetBstMidEndSepPunct{\mcitedefaultmidpunct}
{\mcitedefaultendpunct}{\mcitedefaultseppunct}\relax
\EndOfBibitem
\bibitem[O'Hern \emph{et~al.}(2002)O'Hern, Langer, Liu, and Nagel]{Oh02}
C.~S. O'Hern, S.~A. Langer, A.~J. Liu and S.~R. Nagel, \emph{Phys. Rev. Lett.},
  2002, \textbf{88}, 075507\relax
\mciteBstWouldAddEndPuncttrue
\mciteSetBstMidEndSepPunct{\mcitedefaultmidpunct}
{\mcitedefaultendpunct}{\mcitedefaultseppunct}\relax
\EndOfBibitem
\bibitem[Wilken \emph{et~al.}(2021)Wilken, Guerra, Levine, and Chaikin]{Wi21}
S.~Wilken, R.~E. Guerra, D.~Levine and P.~M. Chaikin, \emph{Phys. Rev. Lett.},
  2021, \textbf{127}, 038002\relax
\mciteBstWouldAddEndPuncttrue
\mciteSetBstMidEndSepPunct{\mcitedefaultmidpunct}
{\mcitedefaultendpunct}{\mcitedefaultseppunct}\relax
\EndOfBibitem
\bibitem[Atkinson \emph{et~al.}(2013)Atkinson, Stillinger, and Torquato]{At13}
S.~Atkinson, F.~H. Stillinger and S.~Torquato, \emph{Phys. Rev. E}, 2013,
  \textbf{88}, 062208\relax
\mciteBstWouldAddEndPuncttrue
\mciteSetBstMidEndSepPunct{\mcitedefaultmidpunct}
{\mcitedefaultendpunct}{\mcitedefaultseppunct}\relax
\EndOfBibitem
\bibitem[Donev \emph{et~al.}(2005)Donev, Torquato, and Stillinger]{Do05c}
A.~Donev, S.~Torquato and F.~H. Stillinger, \emph{Phys. Rev. E}, 2005,
  \textbf{71}, 011105: 1--14\relax
\mciteBstWouldAddEndPuncttrue
\mciteSetBstMidEndSepPunct{\mcitedefaultmidpunct}
{\mcitedefaultendpunct}{\mcitedefaultseppunct}\relax
\EndOfBibitem
\bibitem[Torquato and Stillinger(2010)]{To10c}
S.~Torquato and F.~H. Stillinger, \emph{Rev. Mod. Phys.}, 2010, \textbf{82},
  2633\relax
\mciteBstWouldAddEndPuncttrue
\mciteSetBstMidEndSepPunct{\mcitedefaultmidpunct}
{\mcitedefaultendpunct}{\mcitedefaultseppunct}\relax
\EndOfBibitem
\bibitem[Torquato(2000)]{To00a}
S.~Torquato, \emph{Int. J. Solids Structures}, 2000, \textbf{37},
  411--422\relax
\mciteBstWouldAddEndPuncttrue
\mciteSetBstMidEndSepPunct{\mcitedefaultmidpunct}
{\mcitedefaultendpunct}{\mcitedefaultseppunct}\relax
\EndOfBibitem
\bibitem[Kim and Torquato(2020)]{Ki20a}
J.~Kim and S.~Torquato, \emph{Proc. Nat. Acad. Sci.}, 2020, \textbf{117},
  8764--8774\relax
\mciteBstWouldAddEndPuncttrue
\mciteSetBstMidEndSepPunct{\mcitedefaultmidpunct}
{\mcitedefaultendpunct}{\mcitedefaultseppunct}\relax
\EndOfBibitem
\bibitem[Matheron(1993)]{Ma93}
G.~Matheron, \emph{Cahiers de G{\'e}ostatistique}, 1993, \textbf{107},
  107--113\relax
\mciteBstWouldAddEndPuncttrue
\mciteSetBstMidEndSepPunct{\mcitedefaultmidpunct}
{\mcitedefaultendpunct}{\mcitedefaultseppunct}\relax
\EndOfBibitem
\bibitem[Franz \emph{et~al.}(2017)Franz, Parisi, Sevelev, Urbani, and
  Zamponi]{Fr17b}
S.~Franz, G.~Parisi, M.~Sevelev, P.~Urbani and F.~Zamponi, \emph{Scipost
  Phys.}, 2017, \textbf{2}, 019\relax
\mciteBstWouldAddEndPuncttrue
\mciteSetBstMidEndSepPunct{\mcitedefaultmidpunct}
{\mcitedefaultendpunct}{\mcitedefaultseppunct}\relax
\EndOfBibitem
\bibitem[O'Hern \emph{et~al.}(2003)O'Hern, Silbert, Liu, and Nagel]{Oh03}
C.~S. O'Hern, L.~E. Silbert, A.~J. Liu and S.~R. Nagel, \emph{Phys. Rev. E},
  2003, \textbf{68}, 011306\relax
\mciteBstWouldAddEndPuncttrue
\mciteSetBstMidEndSepPunct{\mcitedefaultmidpunct}
{\mcitedefaultendpunct}{\mcitedefaultseppunct}\relax
\EndOfBibitem
\bibitem[Jin and Yoshino(2021)]{Ji21}
Y.~Jin and H.~Yoshino, \emph{Proc. Natl. Acad. Sci. U.S.A.}, 2021,
  \textbf{118}, e2021794118\relax
\mciteBstWouldAddEndPuncttrue
\mciteSetBstMidEndSepPunct{\mcitedefaultmidpunct}
{\mcitedefaultendpunct}{\mcitedefaultseppunct}\relax
\EndOfBibitem
\bibitem[Nocedal(1980)]{No80}
J.~Nocedal, \emph{Math. Comput.}, 1980, \textbf{35}, 773--782\relax
\mciteBstWouldAddEndPuncttrue
\mciteSetBstMidEndSepPunct{\mcitedefaultmidpunct}
{\mcitedefaultendpunct}{\mcitedefaultseppunct}\relax
\EndOfBibitem
\bibitem[Liu and Nocedal(1989)]{Liu89}
D.~C. Liu and J.~Nocedal, \emph{Math. Program.}, 1989, \textbf{45},
  503--528\relax
\mciteBstWouldAddEndPuncttrue
\mciteSetBstMidEndSepPunct{\mcitedefaultmidpunct}
{\mcitedefaultendpunct}{\mcitedefaultseppunct}\relax
\EndOfBibitem
\bibitem[Newman and Ziff(2000)]{Ne00}
M.~E.~J. Newman and R.~M. Ziff, \emph{Phys. Rev. Lett.}, 2000, \textbf{85},
  4104--4107\relax
\mciteBstWouldAddEndPuncttrue
\mciteSetBstMidEndSepPunct{\mcitedefaultmidpunct}
{\mcitedefaultendpunct}{\mcitedefaultseppunct}\relax
\EndOfBibitem
\bibitem[Vanoni \emph{et~al.}(2025)Vanoni, Kim, Steinhardt, and Torquato]{Va25}
C.~Vanoni, J.~Kim, P.~J. Steinhardt and S.~Torquato, \emph{Dynamical properties
  of particulate composites derived from ultradense stealthy hyperuniform
  sphere packings}, 2025\relax
\mciteBstWouldAddEndPuncttrue
\mciteSetBstMidEndSepPunct{\mcitedefaultmidpunct}
{\mcitedefaultendpunct}{\mcitedefaultseppunct}\relax
\EndOfBibitem
\bibitem[Torquato(2021)]{To21c}
S.~Torquato, \emph{Phys. Rev. E}, 2021, \textbf{103}, 052126\relax
\mciteBstWouldAddEndPuncttrue
\mciteSetBstMidEndSepPunct{\mcitedefaultmidpunct}
{\mcitedefaultendpunct}{\mcitedefaultseppunct}\relax
\EndOfBibitem
\bibitem[Torquato \emph{et~al.}(2000)Torquato, Truskett, and
  Debenedetti]{To00b}
S.~Torquato, T.~M. Truskett and P.~G. Debenedetti, \emph{Phys. Rev. Lett.},
  2000, \textbf{84}, 2064--2067\relax
\mciteBstWouldAddEndPuncttrue
\mciteSetBstMidEndSepPunct{\mcitedefaultmidpunct}
{\mcitedefaultendpunct}{\mcitedefaultseppunct}\relax
\EndOfBibitem
\bibitem[Atkinson \emph{et~al.}(2016)Atkinson, Zhang, Hopkins, and
  Torquato]{At16a}
S.~Atkinson, G.~Zhang, A.~B. Hopkins and S.~Torquato, \emph{Phys. Rev. E},
  2016, \textbf{94}, 012902\relax
\mciteBstWouldAddEndPuncttrue
\mciteSetBstMidEndSepPunct{\mcitedefaultmidpunct}
{\mcitedefaultendpunct}{\mcitedefaultseppunct}\relax
\EndOfBibitem
\end{mcitethebibliography}

\providecommand*{\mcitethebibliography}{\thebibliography}
\csname @ifundefined\endcsname{endmcitethebibliography}
{\let\endmcitethebibliography\endthebibliography}{}

\end{document}